# Utilizing the Hamiltonian dynamics to study resonant interactions of whistler-mode waves and electrons in the solar wind

Tien Vo

May 5, 2021

# Acknowledgements

First and foremost, I thank Professor Cattell for having encouraged me to explore the multiple aspects of this project. With her assistance and advice, I have grown as a researcher from having the freedom to utilize my interests in math, computation, and physics. Second, I thank Professor Lysak for his help in the derivations of the once very intimidating math in this thesis. I would also like to thank Dr. Vadim Roytershteyn for the insightful conversations about PIC simulations, which greatly helped shape my code. I thank Aaron West for his great contributions to the research. And finally, I thank my family for their love and support of my academic pursuit. This thesis owes its existence to their belief in me.

**Abstract**

The role of large amplitude whistler waves in the energization and scattering of solar wind electrons has long been an interesting problem in Space Physics. To study this wave-particle interaction, we developed a vectorized test particle simulation with a variational calculation of the Lyapunov exponents. From using secular perturbation theory on this Hamiltonian system of wave and particle, we confirmed that the pitch angle diffusion of the particle was along the constant Hamiltonian surface and that it was driven by the interaction with the resonance surfaces. We also showed that oblique whistlers could efficiently scatter field-aligned strahl electrons into the halo population in the solar wind. We demonstrated through simulation that these waves were capable of generating horn-like features in the velocity distribution function, similar to recent PIC simulation results in the literature.

# Contents

# 1   Introduction

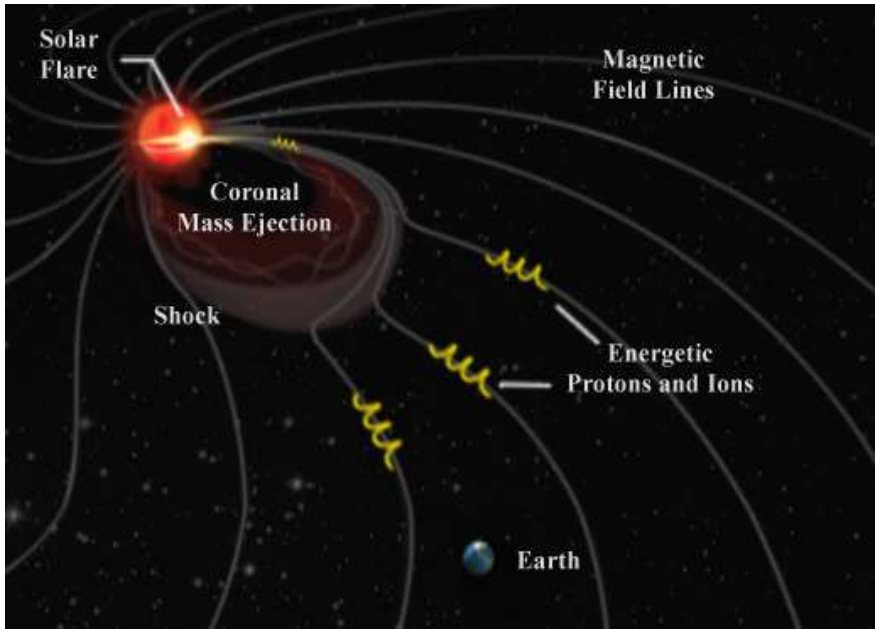

Figure 1: The spiral geometry of the solar wind and the interplanetary magnetic field lines (Mitchell et al., 2019).

The solar wind, being constantly released from the solar corona, is a magnetized and nearly collisionless plasma consisting primarily of electrons, protons, and alpha particles. Typically, it can be described as a magnetohydrodynamic fluid with very high magnetic Reynold's number. Consequently, the magnetic field at the solar surface is frozen into the solar wind plasma and carried along with it. This results in a spiralled geometry of the interplanetary magnetic field lines called the Parker spirals (see Fig. 1). Parker (1958) found from this geometry that the magnetic field followed an inverse square law $B_r \sim r^{-2}$ and the particle density $n \sim r^{-2}V^{-1}$ also decreased with increasing speed $V$ and radial distance.

In the velocity distribution of solar wind electrons, observations have shown that there are usually three populations, a cold core, a hot halo, and a magnetic field aligned strahl, which evolve with heliospheric distance (Montgomery et al., 1968; Feldman et al., 1975; Pilipp et al., 1987). Observations near the Sun (0.3 AU) from the Parker Solar Probe (PSP) have reported that the halo almost disappears, while the strahl is narrower than further out from the Sun (Halekas et al., 2020a). Fig. 2 shows an example of the velocity distribution function (VDF) of electrons at 0.3 AU, where both core and strahl populations are modelled with a bi-Maxwellian distribution. Far from the Sun, statistical studies at 1 AU from Maksimovic et al. (2005) and Wilson III et al. (2019) have modelled core electrons with a bi-Maxwellian distribution, while the halo and the strahl are better fitted with a bi-Kappa distribution (see Fig. 3).

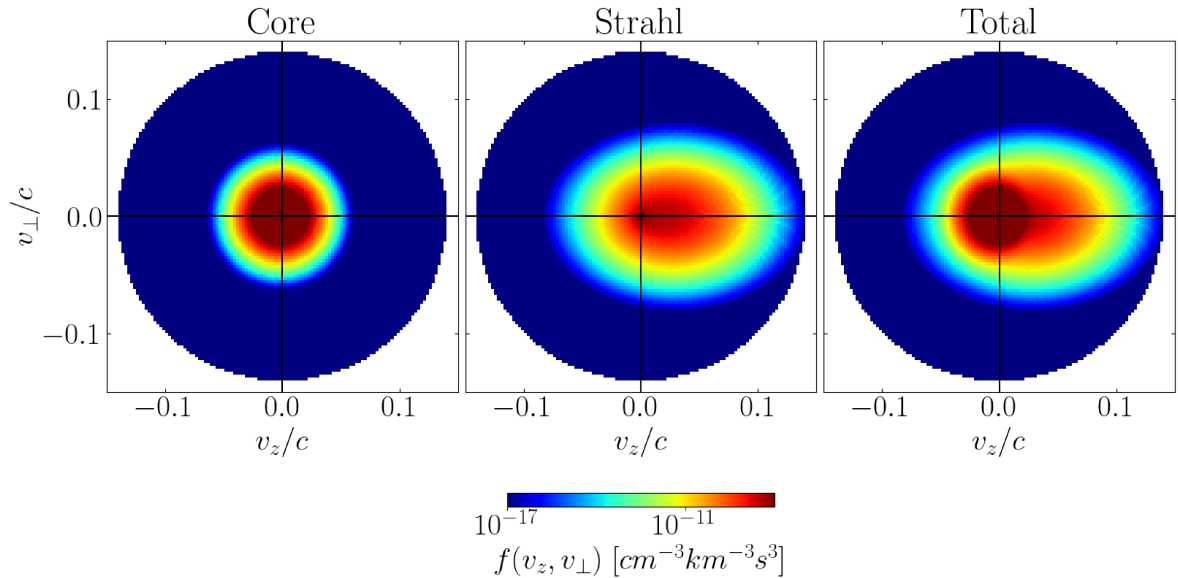

Figure 2: Model of initial electron populations at 0.3 AU used in the simulations in Micera et al. (2020). The core and strahl bulk velocity ensures zero net current.

As these electrons stream radially out, if their propagation were adiabatic, meaning the magnetic moment $\mu \sim v_\perp^2/B_r$ were conserved, then the perpendicular velocity would have to decrease. This means far from the Sun, the strahl should be increasingly narrow. However, in-situ data have shown an opposite trend in the radial evolution of solar wind electrons. Štverák et al. (2009) observed from 0.3 to 1 AU that the strahl density decreased as the halo density increased by the same amount relative to the core (see Fig. 4). This suggests that the origin of the halo is due to the scattering of the strahl. Additionally, Anderson et al.

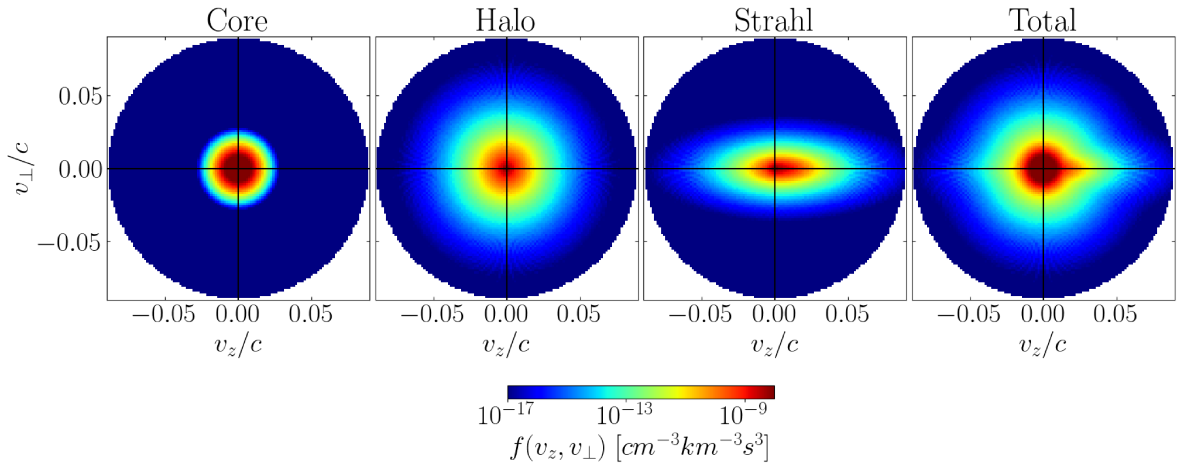

Figure 3: Components of solar wind electrons observed by the Wind satellite at 1 AU as fitted by Wilson III et al. (2019).

(2012) and Graham et al. (2017) reported that the strahl's pitch angle width distribution varied greatly from 5 to 90° at 1 AU and increased radially beyond 1 AU. Thus, it would be harder to identify a field aligned strahl population further out from the Sun.

Therefore, there must be a mechanism that scatters strahl electrons into the halo distribution. Wave-particle interaction is one such process that allows the energization and scattering of resonant electrons. Specifically, whistler-mode waves, a right-hand polarized electromagnetic wave, have long been proposed as a candidate to explain these solar wind observations. Through theoretical and simulation studies, they have been demonstrated to scatter electrons in the Earth's radiation belts (Karimabadi et al., 1990; Albert, 1993; Tao & Bortnik, 2010; Hsieh & Omura, 2017, and references therein). However, these studies typically focused on small whistler amplitudes with $\delta B/B_0 \sim \mathcal{O}(10^{-4})$. Breneman et al. (2010) and Cattell et al. (2020) used electric field waveform captures from the STEREO satellites at 1 AU and demonstrated that large amplitude, narrowband, obliquely propagating whistlers were frequently present in the solar wind. They were observed in the range of 5–40 mV/m, which corresponds to $\delta B/B_0 \sim \mathcal{O}(0.1)$. These large amplitude whistlers recently become an interest because of new data from PSP at 0.3 AU. Agapitov et al. (2020) and Cattell et al. (2021a) observed large amplitude waves of this order near the Sun. Additionally, their polarization indicated that the propagation varied from quasi-parallel to oblique angles. Micera et al. (2020) simulated whistlers from heat-flux instabilities near the Sun using electron distributions modelled after PSP data and showed the halo formation from strahl electrons. Roberg-Clark et al. (2019) reported the formation of "horns" in velocity space due to the scattering of resonant strahl electrons with oblique whistlers in solar flares (see Fig. 5). Thus, we are interested in studying the scattering and energization of solar wind electrons due to these large amplitude waves and comparing our results with observations and these recent simulations.

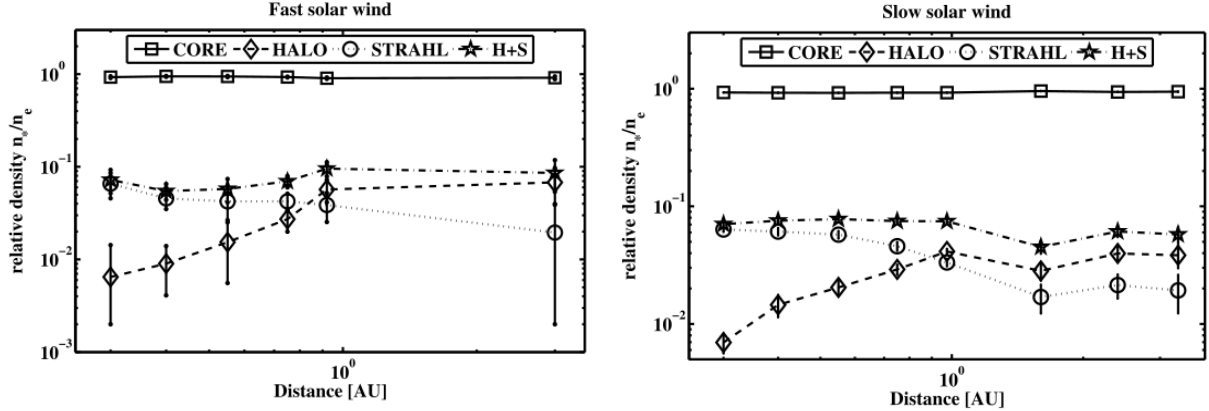

Figure 4: Radial evolution of electrons in the fast and slow solar wind from 0.3 to 6 AU (Štverák et al., 2009).

Kersten (2014) developed a test particle simulation to study whistler-electron interactions in the radiation belts and later adapted it to simulate whistlers at stream interaction regions in the solar wind based on observations in Breneman et al. (2010). Modelled after the simulation in Roth et al. (1999), the code used a fourth order Runge-Kutta (RK) integration algorithm to solve the Lorentz equation numerically. This is a general approach to numerical problems, as the RK family of integrators is known to produce highly precise solutions. The results are therefore reliable as long as one is interested in single-particle behaviors. However, this approach fails to maintain the consistency among a spectrum of initial conditions as the solutions might be more unstable for certain regions in phase space. For the high amplitude waves of interest, chaotic behavior is usually present. Thus, this program is insufficient to investigate the behavior of an electron distribution as it provides no physical measure to judge the reliability among the results.

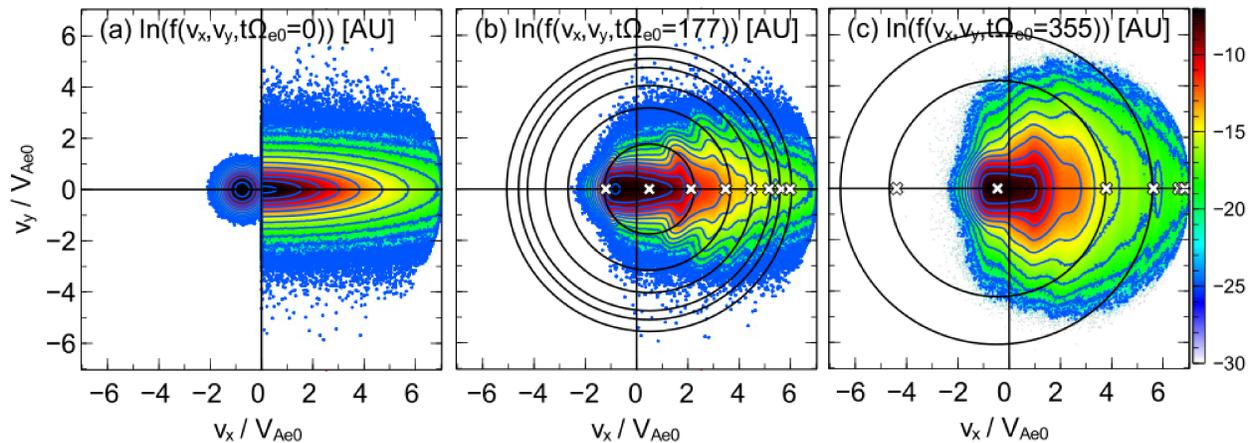

Figure 5: Formation of horns in velocity space of an original distribution (a) of core and strahl electrons (Roberg-Clark et al., 2019). The horizontal and vertical axes are the parallel and perpendicular velocity normalized by the electron Alfvén speed. Panel (b) shows the resulting interaction with a right-ward wave. The white crosses are the $n = -5, -4, ..., 1$ resonances. Panel (c) shows that with a left-ward wave (with $n = -1, 0, ..., 5$ resonances).

Particle-In-Cell (PIC) simulations, as used by Micera et al. (2020) and Roberg-Clark et al. (2019), are a standard in plasma research in studying self-consistently evolving systems. Instead of a high order RK algorithm, they usually use a time-centered, second-order explicit integrator called the Boris pusher (Birdsall & Langdon, 1985). Although not a symplectic algorithm, it is the *de facto* method for advancing a charged particle in an electromagnetic field because the Boris pusher conserves local phase space volume (Qin et al., 2013). This means the energy error is globally bounded for an arbitrarily large number of time steps. Thus, this numerical method is capable of resolving multi-scale dynamical problems over a long simulation period. However, PIC simulations are computationally expensive, as they solve Maxwell equations along with advancing particles and usually handle millions to trillions of particles. For our purpose, large scale PIC simulations are not necessary, because test particle simulations allow us to examine the interaction for different wave properties and over all particle angles and energies.

In this thesis, we use a vectorized test particle simulation capable of investigating the behavior of a distribution of hundreds of thousands of electrons. The code is modelled after the Vector Particle-In-Cell (VPIC) code using only the particle advancing component (Bowers et al., 2008). In Section 2, we derive the whistler wave fields from a cold, collisionless plasma dispersion relation and also establish the Hamiltonian analysis of the resonance surface using Hamilton-Jacobi and perturbation theory. In Section 3, we lay out the detail of the calculations in the simulation and discuss the estimation of the Lyapunov exponents to measure the efficiency of the integration algorithm. In Section 4, we present the diagnostics of the simulation including the Lyapunov exponents, the adiabatic invariants, and whistler parameters. In Section 5, we present simulation results of the electron distribution interactions with single uniform whistlers and a narrowband packet of whistlers at 0.3 AU and 1 AU and the analysis of these results as according to quasi-linear resonant theory. Conclusions and suggestions for future works are in Sections 6 and 7.

# 2 Theory

## 2.1 Equations of whistler wave fields

In a cold uniform plasma with a background magnetic field $\mathbf{B}_0 = B_0\hat{\mathbf{z}}$, the electric permittivity tensor is

$$\boldsymbol{\epsilon} = \epsilon_0\boldsymbol{\epsilon}_R = \epsilon_0 \begin{pmatrix} S & -iD & 0 \\ iD & S & 0 \\ 0 & 0 & P \end{pmatrix} \qquad (2.1)$$

where the constants $S, D, P$ are the Stix parameters (Stix, 1992) given as follows

$$S = 1 - \sum_s \frac{\omega_{ps}^2}{\omega^2 - \Omega_{cs}^2}, \qquad D = \sum_s \frac{|q_s|}{q_s} \frac{\Omega_{cs} \omega_{ps}^2}{\omega(\omega^2 - \Omega_{cs}^2)}, \qquad P = 1 - \sum_s \frac{\omega_{ps}^2}{\omega^2} \qquad (2.2)$$

The summation is over all species $s$ with charge $q_s$, mass $m_s$, and density $n_s$. The plasma frequency is $\omega_{ps} = \sqrt{n_s q_s^2 / \epsilon_0 m_s}$, and $\Omega_{cs} = |q_s| B_0 / m_s$ is the cyclotron frequency. Now, let there be an electromagnetic wave propagating in the $(xz)$ plane with $\mathbf{k} = k_\perp \hat{\mathbf{x}} + k_\parallel \hat{\mathbf{z}} = k(\sin\theta \hat{\mathbf{x}} + \cos\theta \hat{\mathbf{z}})$. Assume also that the fields are Fourier transformed so that $\boldsymbol{\nabla} \rightarrow i\mathbf{k}$ and $\partial/\partial t \rightarrow -i\omega$. From Maxwell equations, the electric field satisfies $\mathbf{N} \times (\mathbf{N} \times \mathbf{E}) + \boldsymbol{\epsilon}_R \cdot \mathbf{E} = 0$ where $\mathbf{N} = c\mathbf{k}/\omega$ is the refractive index. This can be written in the form $\mathbf{R} \cdot \mathbf{E} = 0$ where

$$\det \mathbf{R} = \det \begin{pmatrix} S - N_\parallel^2 & -iD & N_\perp N_\parallel \\ iD & S - N^2 & 0 \\ N_\perp N_\parallel & 0 & P - N_\perp^2 \end{pmatrix} = 0 \qquad (2.3)$$

from which the refractive index can be solved. Plugging it back into $\mathbf{R} \cdot \mathbf{E} = 0$ yields the electric field polarizations. The right-hand polarized solution with frequencies between $\Omega_{ci}$ and $\Omega_{ce}$ is called the whistler mode whose fields can be written in the form

$$\mathbf{B}_w = B_x^w \sin\psi \hat{\mathbf{x}} + B_y^w \cos\psi \hat{\mathbf{y}} + B_z^w \sin\psi \hat{\mathbf{z}} \qquad (2.4a)$$

$$\mathbf{E}_w = E_x^w \cos\psi \hat{\mathbf{x}} - E_y^w \sin\psi \hat{\mathbf{y}} + E_z^w \cos\psi \hat{\mathbf{z}} \qquad (2.4b)$$

where the wave phase is $\psi = \mathbf{k} \cdot \mathbf{r} - \omega t$ and the magnetic field is given by Faraday's law $\mathbf{B}_w = (1/\omega)\mathbf{k} \times \mathbf{E}_w$. The polarizations are summarized in Tao & Bortnik (2010)

$$E_x^w / E_x^w = 1 \qquad E_y^w / E_x^w = \frac{D}{N^2 - S} \qquad E_z^w / E_x^w = -\frac{N^2 \sin\theta \cos\theta}{P - N^2 \sin^2\theta}$$

$$cB_x^w / E_x^w = \frac{ND\cos\theta}{N^2 - S} \qquad cB_y^w / E_x^w = \frac{NP\cos\theta}{P - N^2 \sin^2\theta} \qquad cB_z^w / E_x^w = \frac{ND\sin\theta}{S - N^2} \qquad (2.5)$$

For the analysis of the Hamiltonian, it is also necessary to find a scalar and vector potential representing the above fields. Assuming the general form for the whistler potentials used in Karimabadi et al. (1990) and Roth et al. (1999), we can find the amplitudes such that they are consistent with Eq. (2.4). Suppose the scalar potential is $\Phi_w = \Phi_0 \sin \psi$ and the vector potential is

$$\mathbf{A}_w = A_1 \left( \frac{k_\parallel}{k} \right) \sin \psi \hat{\mathbf{x}} + A_2 \cos \psi \hat{\mathbf{y}} - A_1 \left( \frac{k_\perp}{k} \right) \sin \psi \hat{\mathbf{z}} \tag{2.6}$$

Equating the corresponding electric field $\mathbf{E} = -\boldsymbol{\nabla} \Phi_w - \partial \mathbf{A}_w / \partial t$ to Eq. (2.4b), we can solve for $\Phi_0, A_1$, and $A_2$ as follows.

$$\Phi_0 = -\frac{1}{k} \left[ \left( \frac{k_\perp}{k} \right) E_x^w + \left( \frac{k_\parallel}{k} \right) E_z^w \right] \quad A_1 = \frac{1}{\omega} \left[ \left( \frac{k_\parallel}{k} \right) E_x^w - \left( \frac{k_\perp}{k} \right) E_z^w \right] \quad A_2 = \frac{E_y^w}{\omega} \tag{2.7}$$

## 2.2 Particle dynamics

The curvature of the Parker spiral is small over a length scale of $\sim 100\,000$ km, which we will later confirmed through comparison with the particle motion. We can therefore assume the background field is uniform $\mathbf{B}_0 = B_0 \hat{\mathbf{z}}$. Given a vector potential $\mathbf{A} = \mathbf{A}_w + x B_0 \hat{\mathbf{y}}$ and a scalar potential $\Phi_w$ where $\mathbf{A}_w, \Phi_w$ are defined as in Section 2.1, the relativistic Hamiltonian for a particle with mass $m$ and charge $q$, is

$$\mathcal{H} = \sqrt{m^2 c^4 + (\mathbf{P} - q\mathbf{A}_w - q B_0 x \hat{\mathbf{y}})^2 c^2} + q\Phi_w \tag{2.8}$$

where $\mathbf{P} = \gamma m \mathbf{v} + q\mathbf{A}$ is the canonical momentum conjugate to the Cartesian coordinates.

There are two issues. First, note that $\mathcal{H}$ depends on $x$, so $\dot{P}_x = -\partial \mathcal{H} / \partial x \neq 0$ and $P_x$ is not invariant. Secondly, $\mathbf{A}_w$ oscillates with the phase $\psi(x, z, t)$. So the energy is not conserved as the Hamiltonian is time-dependent. The former is a standard problem since $\mathcal{H}$ is currently formulated in Cartesian coordinates, whereas the system is cylindrically symmetric due to the background magnetic field. This can be resolved by transforming into a cylindrical frame (Goldstein et al., 2002). The latter is, however, more problematic as the wave introduces oscillations symmetric about its direction of propagation. In-depth analysis of the Hamiltonian can be done by using secular perturbation theory (Lichtenberg & Lieberman, 1992), which involves decomposing the Hamiltonian into Bessel-Fourier series and performing the gyro-averaging method to separate a single term, the $n$th harmonic, in the series.

Within the scope of our analysis, we will calculate this Hamiltonian system's adiabatic invariants and derive its resonance surfaces similar to the approach of Karimabadi et al. (1990) and Roberg-Clark et al. (2019). The mathematical details are given in Appendix B. For motion near the resonance $n$, the Hamiltonian can be recast into the form

$$\mathcal{H}(\zeta; \hat{P}_\phi, \hat{P}_\zeta) = \gamma\left(\hat{P}_\phi + n\hat{P}_\zeta, k_\parallel \hat{P}_\zeta\right) mc^2 - \omega\hat{P}_\zeta + G_n\left(\hat{P}_\phi + n\hat{P}_\zeta, k_\parallel \hat{P}_\zeta\right)\sin\zeta \quad (2.9)$$

where the action-angle variables $(\zeta, \hat{P}_\zeta)$ and $(\phi, \hat{P}_\phi)$ are given by

$$\zeta = n\phi + k_\perp P_y/qB_0 + k_\parallel z - \omega t \qquad \phi = \tan^{-1}\left[\frac{m\Omega_c\left(x - P_y/qB_0\right)}{P_x}\right]$$

$$\hat{P}_\zeta = P_\parallel/k_\parallel \qquad\qquad \hat{P}_\phi = P_\phi - nP_\parallel/k_\parallel = P_\perp^2/2m\Omega_c - nP_\parallel/k_\parallel \quad (2.10)$$

The perpendicular momentum is defined as $P_\perp = \sqrt{P_x^2 + \left(P_y - qB_0 x\right)^2}$. The gyroradius is then $\rho = P_\perp/m\Omega_c = \sqrt{2P_\phi}$, and $\gamma = \sqrt{1 + (P_\perp^2/m^2c^2) + (P_\parallel^2/m^2c^2)}$ is the Lorentz factor. The perturbation amplitude $G_n$ is defined as

$$G_n(P_\phi, P_\parallel) = mc^2\left\{s\left[\delta_0 + \frac{\delta_1}{\gamma}\left(\frac{k_\perp}{k}\frac{P_\parallel}{mc} - \frac{k_\parallel}{k}\frac{n\Omega_c}{ck_\perp}\right)\right]J_n\left(k_\perp\sqrt{2P_\phi}\right) + \frac{\delta_2}{\gamma}\frac{\rho\Omega_c}{c}J_n'\left(k_\perp\sqrt{2P_\phi}\right)\right\}$$
$$(2.11)$$

where $J_n, J_n'$ are the $n$th order Bessel functions of the first kind and their derivatives, the wave potential amplitudes are $\delta_0 = |q|\Phi_0/mc^2$ and $\delta_{1,2} = |q|A_{1,2}/mc$, and $s = q/|q|$ is the charge sign. The equation of motion of this system is

$$\frac{d\zeta}{dt} = -\omega + \frac{n\Omega_c}{\gamma} + \frac{k_\parallel P_\parallel}{\gamma m} + \left( n\frac{\partial G_n}{\partial P_\phi} + k_\parallel \frac{\partial G_n}{\partial P_\parallel} \right) \sin\zeta \tag{2.12a}$$

$$\frac{d\hat{P}_\zeta}{dt} = -G_n \cos\zeta \tag{2.12b}$$

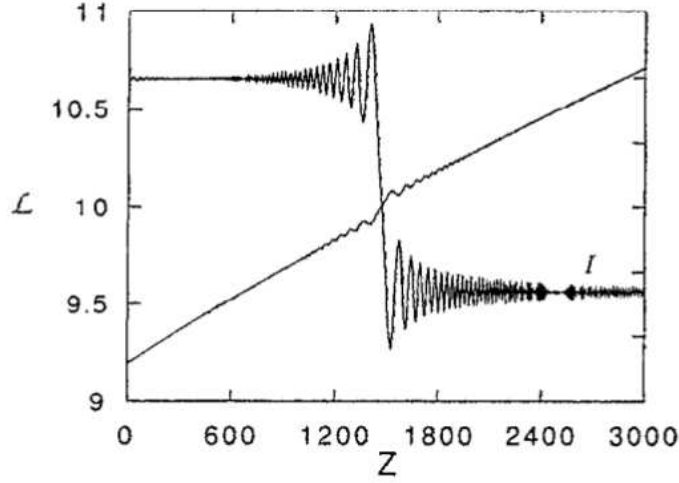

Figure 6: The change in the adiabatic invariant $I$ as the resonance $\mathcal{L}$ crosses an integer value (figure from Albert (1993)). In our notations, $\mathcal{L} = n$, $I = \hat{P}_\phi$, and $Z = z$.

Here, we have assumed that the wave is small ($\delta_{0,1,2} \ll 1$ and $\delta_{1,2} < \gamma v/c$, where $v$ is the particle's velocity). So the motion $\dot{\zeta}$ is usually fast, meaning we can average over $\zeta$ and $\dot{P}_\zeta = 0$, except for when

$$\omega = \frac{n\Omega_c}{\gamma} + \frac{k_\parallel P_\parallel}{\gamma m} \tag{2.13}$$

The adiabatic invariant $\hat{P}_\zeta$ is no longer conserved whenever the particle undergoes a resonance crossing (see Fig. 6). Eq. (2.13) then describes a resonant condition. Although this is not a convention, most papers in the literature define the gyrophase as $s\phi$, which results in the resonant mode being $sn$. For an electron with $s = -1$, this means their fundamental cyclotron motion is the $n = -1$ mode, while our fundamental cyclotron as defined by

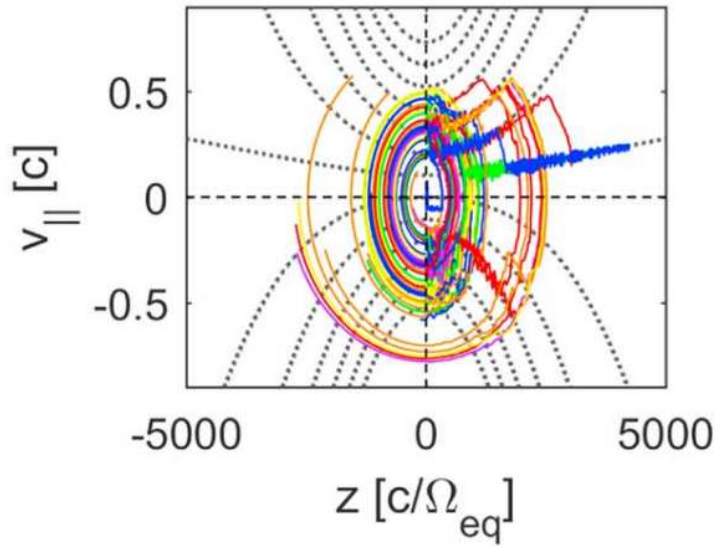

Figure 7: Trajectories (colored solid lines) of particles being trapped along (dotted) resonance lines ($|n| \leq 6$) in Hsieh & Omura (2017).

Eq. (2.13) is $n = 1$. Since nothing changes but the naming, we shall use our own definitions in this thesis. More physics can be described from here, including the characterization of resonant responses. Particles can either be scattered or trapped into resonance (see Fig. 7). It involves expanding the Hamiltonian around the resonances and investigating the separatrices in phase space (Karimabadi et al., 1990; Artemyev et al., 2018). This is outside the scope of this thesis.

## 2.3 Resonance surfaces

Using the dynamics we established in Section 2.2, we can use a tool provided by Karimabadi et al. (1990), the resonance-diagram technique. The derivation steps are included in Appendix C. Let $H_0 = \gamma - v_p(P_\parallel/mc)$ be the normalized unperturbed Hamiltonian in Eq. (2.9) where $v_p = 1/N_\parallel = \omega/k_\parallel c$ is the normalized phase velocity and $N_\parallel$ the parallel refractive index. A constant value of $H_0$ defines a constant energy ($H$) surface in phase space ($P_\perp, P_\parallel$). In the non-relativistic limit, this is the equation of a circle centered around $v_p$ with $(v_\parallel/c - v_p)^2 + v_\perp^2/c^2 = $ const (Roberg-Clark et al., 2019). In the relativistic limit, the $H$ surface is elliptic

$$\frac{(v_\parallel - v_c)^2/c^2}{R_0/(H_0^2 + v_p^2)} + \frac{v_\perp^2/c^2}{R_0/H_0^2} = 1 \tag{2.14}$$

where $v_c/c = v_p/(H_0^2 + v_p^2)$ and $R_0 = H_0^2 - 1 + v_p^2/(H_0^2 + v_p^2)$. We have approximated $P \approx \gamma m v$ (which is valid if the particle term dominates in the canonical momentum) and write the surface in terms of the observable $v$. Similarly, one can also define a resonance ($R$) surface from the resonant condition Eq. (2.13). Its intersections to the $v_\perp = 0$ axis are

$$v_{r,\parallel} = \frac{v_p}{1 + \alpha_n^2} \pm \sqrt{\frac{\alpha_n^2}{1 + \alpha_n^2}\left(1 - \frac{v_p^2}{1 + \alpha_n^2}\right)} \tag{2.15}$$

where $\alpha_n = n\Omega_c/k_\parallel c$. The Landau resonance ($n = 0$) is located at the center of all $H$ surfaces and other pairs of resonance ($n = \pm 1, \pm 2, \pm 3, \ldots$) are equidistant to that center (see Fig. 5 for examples from the Roberg-Clark simulation). For whistler waves, $N_\parallel$ is usually larger than 1, so the maximum energization is highly limited because the number of $H$–$R$ intersections are small (Karimabadi et al., 1990). Thus, particles tend to move along the constant $H$ surface until they interact resonantly with the wave near the $H$–$R$ intersection and become energized or de-energized. In subsequent sections, we will only investigate particles in the non-relativistic energy range where $H$ is circular and $R$ is approximately a constant surface at $v_z = v_{r,\parallel}$. In our analysis, we will confirm that the particles' trajectories in phase space follow this behavior.

# 3  Simulation

## 3.1  Particle advance

The Hamiltonian equation of motion in Eq. (2.12), although useful for analysis, is only an approximation near a single resonance. Roth et al. (1999) alternated between that and the exact Lorentz force to reduce the computational cost for particles entering resonance, since adaptive RK of the 4th order is expensive. However, in doing so, the code user must impose an arbitrary boundary in switching between the resonant and non-resonant regimes. Here, we shall use the relativistic Boris pusher from Ripperda et al. (2018) to solve for the full Lorentz force and rely on its volume-preserving characteristics to choose the appropriate step size. However, we must first describe our normalizations. From Section 2.1, it is natural to normalize $\mathbf{B} \to \mathbf{B}/B_0$ and subsequently $\mathbf{E} \to \mathbf{E}/cB_0$. Since we are using relativistic formulations, $\mathbf{v} \to \mathbf{v}/c$ and $\mathbf{P} \to \mathbf{P}/mc$. The characteristic frequency in our system is defined by the electron cyclotron frequency $\Omega_{ce}$, so the wave frequency $\omega \to \omega/\Omega_{ce}$ and time $t \to t\Omega_{ce}$. The spatial position thus becomes $\mathbf{r} \to \mathbf{r}\Omega_{ce}/c$.

The description of the Boris algorithm is as follows. The Lorentz force in natural units has the form $d\mathbf{u}/dt = s(\mathbf{E} + \mathbf{v} \times \mathbf{B})$ where $\mathbf{u} = \gamma\mathbf{v}$ and $\gamma = \sqrt{1 + u^2}$. The time-centered finite difference expression of this is

$$\mathbf{u}_{n+1} - \mathbf{u}_n = s\Delta t\Big[\mathbf{E}_n + \big(1/2\gamma_n\big)(\mathbf{u}_{n+1} + \mathbf{u}_n) \times \mathbf{B}_n\Big] \tag{3.1}$$

where $\mathbf{u}_n = \gamma_n\mathbf{v}\big(t_n - \Delta t/2\big)$, $\mathbf{E}_n = \mathbf{E}_n(t_n)$, $\mathbf{B}_n = \mathbf{B}_n(t_n)$ and $\Delta t$ is the step size where $t_n = n\Delta t$ for $n \in \mathbb{N}$. $\gamma_n$ is the Lorentz factor determined from $u_n$. Now, the Kick-Drift-Kick steps that make this algorithm a leapfrog scheme are defined via the two auxilliary vectors $\mathbf{u}_\pm$. The first kick is a half electric field acceleration from $\mathbf{u}_n$ to $\mathbf{u}_- = \mathbf{u}_n + (s\Delta t/2)\mathbf{E}_n$, followed by a rotation $\mathbf{u}_- \to \mathbf{u}_+$ by the magnetic field $\mathbf{u}_+ = \mathbf{u}_- + \big(\Delta t/2\gamma_n\big)(\mathbf{u}_+ + \mathbf{u}_-) \times \mathbf{B}_n$. $\mathbf{u}_+$ here seems to be implicitly defined, but from the geometry of this rotation, it can be computed explicity as $\mathbf{u}_+ = \mathbf{u}_- + (\mathbf{u}_- + \mathbf{u}_- \times \mathbf{T}) \times \mathbf{S}$ with $\mathbf{T} = \big(s\Delta t/2\gamma_n\big)\mathbf{B}_n$ and $\mathbf{S} = 2\mathbf{T}/(1+T^2)$ (see more details in Birdsall & Langdon (1985)). Then the second kick accelerates the particle to the next state $\mathbf{u}_{n+1} = \mathbf{u}_+ + \big(s\Delta t/2\big)\mathbf{E}_n$.

To simulate a single uniform whistler fields in natural units, we can factor out from

Eq. (2.4) that

$$\frac{\mathbf{B}_w}{B_0} = \frac{E_x^w}{cB_0} \left[ \left( \frac{cB_x^w}{E_x^w} \right) \sin\psi \hat{\mathbf{x}} + \left( \frac{cB_y^w}{E_x^w} \right) \cos\psi \hat{\mathbf{y}} + \left( \frac{cB_z^w}{E_x^w} \right) \sin\psi \hat{\mathbf{z}} \right] \qquad (3.2)$$

and similarly,

$$\frac{\mathbf{E}_w}{cB_0} = \frac{E_x^w}{cB_0} \left[ \cos\psi \hat{\mathbf{x}} - \left( \frac{E_y^w}{E_x^w} \right) \sin\psi \hat{\mathbf{y}} + \left( \frac{E_z^w}{E_x^w} \right) \cos\psi \hat{\mathbf{z}} \right] \qquad (3.3)$$

Since the STEREO spacecraft only measured the whistler electric field amplitudes (Breneman et al., 2010), we are using $E_x^w$ as the scaling factor. The unitless polarizations can be computed with Eq. (2.5). Note that the wave phase in natural units is $\psi = \omega(N_\perp x + N_\parallel z - t)$, and that it is zero for particles starting out at the origin at $t = 0$. So originally, the wave has an amplitude $E_w^0 = E_x^w \sqrt{1 + \left( E_z^w / E_x^w \right)^2}$. So we shall choose $E_x^w$ such that $E_w^0$ has a desired physical value. To simulate a wave packet with the same original wave amplitude $E_w^0$ and $N$ frequencies $\omega_j = \omega_1 + (j-1)\Delta\omega$ with spacing $\Delta\omega$, we simply have to repeat the calculations Eq. (3.2) and Eq. (3.3) and write the total fields as $\mathbf{E}_w = \sum_{j=1}^N \mathbf{E}_{w,j}$ and $\mathbf{B}_w = \sum_{j=1}^N \mathbf{B}_{w,j}$.

With these calculations, a description of the particle advance at each time step is completed. The scaling factor is calculated at the beginning of the simulation. So each loop involves (a) calculating new wave phase, constructing the total field, and advancing the particle, (b) the diagnostics, and (c) writing to database. (b) and (c) can be activated at different time intervals.

## 3.2 Estimation of the Lyapunov exponents

As mentioned in the previous section, the Boris pusher guarantees a volume-preserving characteristic. To verify that our simulation's step size is sufficiently small that the algorithm efficiently preserves volume in phase space, we employ a concept from chaos theory called the Lyapunov exponents (Ott, 2002). These exponents essentially describe how a basis spanning a $k$-dimensional space changes under subsequent transformations. For simplicity, suppose we have a 1-D trajectory. If the Lyapunov exponent is $\lambda = 0$, then the space (distance, in this case) around it evolves as $\exp(n\lambda) = 1$ and doesn't contract or expand after $n$ time steps. If $\lambda < 0$, the space eventually reduces to a singular point. This is called an attractor where all trajectories starting out near this one being considered converges. If $\lambda > 0$, all trajectories originally close together eventually diverge and become increasingly far from each other. In higher dimensions, we can describe these distortions through the basis elements that span the phase space (see Fig. 8).

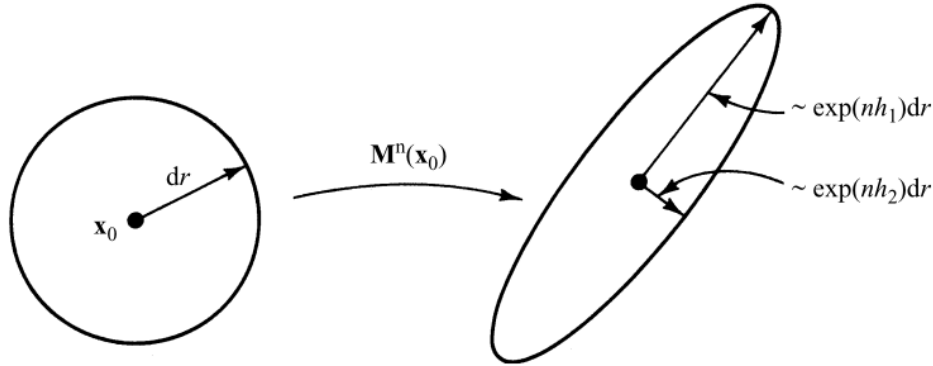

Figure 8: Distortion of a two-dimensional ball after $n$ time steps. $h_1, h_2$ are the Lyapunov exponents in each axial direction spanning the ball.

It requires infinitely many vectors near a point in phase space to compute the Lyapunov exponents precisely. Thus, one can only estimate the values using a variety of methods. Here, we shall use a variational approach with Gram-Schmidt orthogonalization (Benettin et al., 1980; Sandri, 1996). Given an initial condition to our ordinary differential equations (ODEs) in the previous section, we can attach to it a six-dimensional "ball" given by a 6x6 matrix, or a set of six 6-D column vectors $U_0 = \left\{ \mathbf{u}_j \right\}_{j=1}^6$. This choice of a 6-ball is arbitrary, but the 6-D identity map $\mathbb{1}_6$ is an obvious option. It becomes $U_1 = \mathbf{M}_0 \cdot U_0$ after a local expansion $\mathbf{M}_0 = \mathbb{1}_6 + \Delta t \boldsymbol{\nabla}\mathbf{F}_0$ where $\Delta t$ is the step size and $\boldsymbol{\nabla}\mathbf{F}_0$ is the Jacobian of our ODEs at $n = 0$. More details about $\mathbf{M}$ and $\boldsymbol{\nabla}\mathbf{F}$ can be found in Appendix D. By the Gram-Schmidt procedure, we can find a 6-D orthogonal basis $W_1 = \left\{\mathbf{w}_j\right\}_{j=1}^6$ from $U_1$. The volume of the parallelpiped spanned by this new basis is $V_1(W_1) = \prod_{j=1}^6 \|\mathbf{w}_j\|$. Now, the definition of the largest Lyapunov exponent (LCE) after time $t$ is $\lambda = \lim_{t\to\infty}(1/t)\ln V$ where $V$ is the current volume of the 6-ball. So after $N$ time steps, the LCE can be approximated as

$$\lambda = \frac{1}{N\Delta t}\sum_{n=1}^{N}\sum_{j=1}^{6}\ln\left\|\mathbf{w}_j^n\right\| \tag{3.4}$$

where $t \to N\Delta t$ and $\mathbf{w}_j^n$ are the basis elements $j$ at time step $n$. Note the volume is accumulative through time. It is also possible to define separately the Lyapunov exponent in each dimension of the original 6-ball

$$\lambda_j = \frac{1}{N\Delta t}\sum_{n=1}^{N}\ln\left\|\mathbf{w}_j^n\right\| \tag{3.5}$$

Then the LCE is just the sum of $\lambda_j$ over 6 dimensions. Our calculations thus involve consecutively computing at each step $n$ the volume of the ball from $W_n$ and then renormalizing it to measure the expansion of the next advance. The final result is an accumulation of the volume expansion through $N$ time steps, from which the LCE can be calculated.

# 4 Diagnostics

## 4.1 Wave parameters

In subsequent sections, we will study the interactions of whistlers with electrons in two sets of background parameters. The first one is typical of 1 AU with a background field strength $B_0 = 10\,\mathrm{nT}$. The plasma is quasineutral with $n = n_i = n_e = 5\,\mathrm{cm}^{-3}$. The second is consistent with the simulation at 0.3 AU in Micera et al. (2020) with $B_0 = 50\,\mathrm{nT}$ and $n = n_i = n_e = 300\,\text{cm}^{-3}$. The whistler parameters are based on those of Cattell et al. (2020). For both sets of background parameters, the single waves have an amplitude $E_w^0 = 20\,\text{mV/m}$, frequency $\omega/\Omega_{ce} = 0.15$, and propagation angles $\theta = 5$, $65°$, and $175°$. Whistler packets will contain a set of eleven $20\,\text{mV/m}$ single whistlers with frequency from $0.135$ to $0.165\ \Omega_{ce}$ and propagation angles $\theta = 0$, $65°$, and $180°$.

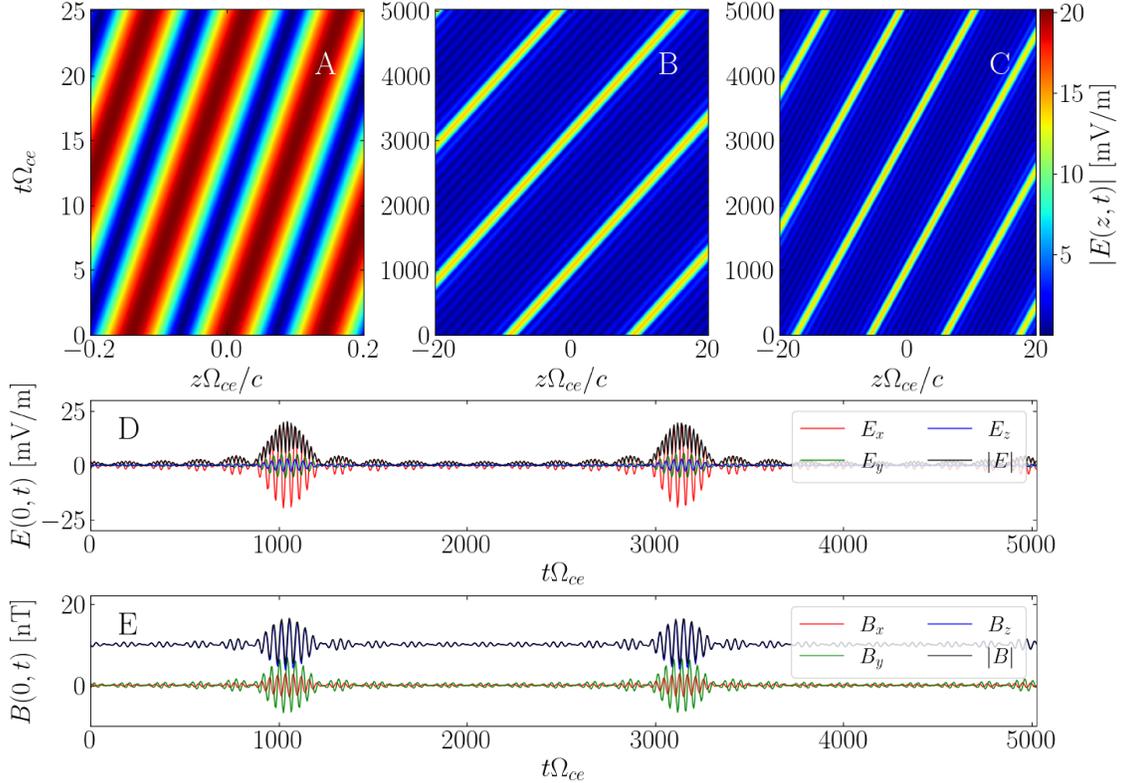

Figure 9: The spatiotemporal evolution at $x = y = 0$ of the electric field of an oblique ($\theta = 65°$) single whistler (A) and oblique whistler packets (B & C). Panels A and B have background parameters for 1 AU and panel C is for 0.3 AU. Panels D and E show the electric and magnetic field components at $z = 0$ of the packet in panel B.

A few examples showing the oblique wavefronts are shown in panels A, B, and C of Fig. 9. The phase velocities are different between 0.3 and 1 AU because the background parameters are different. Panels D and E show in more detail the oblique packet in panel B as observed at the origin in time. In Cattell et al. (2020), the mean observed amplitude was $10\,\mathrm{mV/m}$, while those as high as $40\,\mathrm{mV/m}$ were also observed. Thus, in this study, we use $20\,\mathrm{mV/m}$ which has $\delta B_w/B_0 \sim 0.6$ to clearly see the possible role of the waves. These large amplitude oscillations can result in highly chaotic behavior in the particle motion. Also, note that we are greatly overestimating the parallel wave amplitudes at 1 AU for the sake of comparison. In reality, parallel whistlers at 1 AU are only observed with $\delta B_w/B_0 \sim \mathcal{O}(0.01)$.

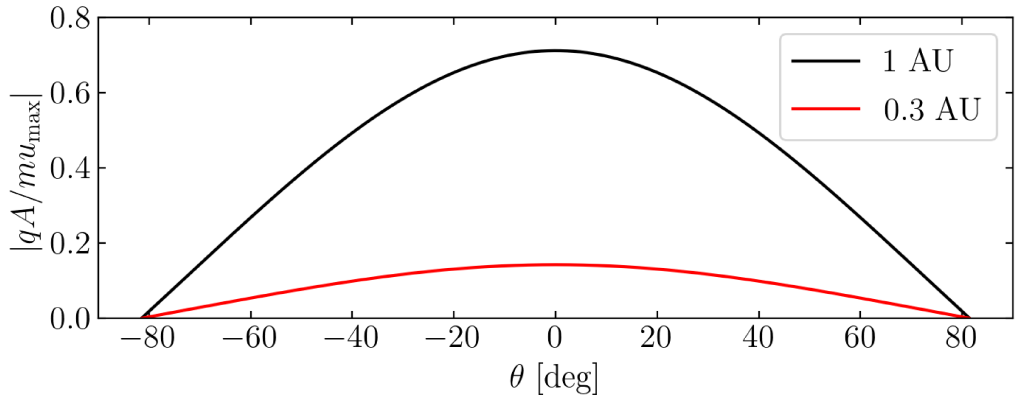

Figure 10: The ratio between the potential field term $qA$ and the particle's relativistic speed term $mu = \gamma m v$ in the canonical momentum. $u_{max}$ corresponds to the maximum kinetic energy of the simulated electrons. Note that whistlers do not propagate beyond the resonance cone angle, which is close to $81°$ but not identical between the black and red curves (Remya et al., 2016).

Since the thermal velocity of electrons in the solar wind is $\sim 2\,000 - 5\,000\,\mathrm{km/s}$, they are fairly non-relativistic (see Fig. 3). Based on observations of the energy range for solar wind electrons, we will initiate particles up to $\sim 1\,\mathrm{keV}$ in our simulations. Fig. 10 verifies the small field assumption $\delta_{1,2} < \gamma v/c$ in Section 2.2 for the resonant condition, with $u_{max}$ corresponding to $1\,\mathrm{keV}$. Consequently, $\delta_{1,2} \sim \mathcal{O}(0.01)$ are small compared to unity. So our assumptions regarding the Hamiltonian derivations are justified, even in the perturbation of these large amplitude whistlers. For this energy range, the maximum $z$ in our simulations is $\sim 30\,000\,\mathrm{km}$, which justifies the assumption of uniform background magnetic field.

## 4.2 Particle parameters

As mentioned in Section 1, the two standard velocity distribution functions (VDF) used to model solar wind electrons are the bi-Maxwellian and the bi-Kappa. The former is

$$f_M(v_\perp, v_\parallel) = \frac{n_0}{\pi^{3/2} v_{th,\perp}^2 v_{th,\parallel}} \exp\left\{ \left[ \left( \frac{v_\parallel - v_{o,\parallel}}{v_{th,\parallel}} \right)^2 + \left( \frac{v_\perp - v_{o,\perp}}{v_{th,\perp}} \right)^2 \right] \right\} \qquad (4.1)$$

where $v_\parallel = v_z$, $v_\perp = \sqrt{v_x^2 + v_y^2}$, $v_{th,j}$ is the thermal speed, $v_{o,j}$ is the drift speed in each direction, and $n_0$ is the population density. The bi-Kappa VDF is given by

$$f_K(v_\perp, v_\parallel) = A_\kappa \left\{ 1 + \left( \kappa - \frac{3}{2} \right)^{-1} \left[ \left( \frac{v_\parallel - v_{o,\parallel}}{v_{th,\parallel}} \right)^2 + \left( \frac{v_\perp - v_{o,\perp}}{v_{th,\perp}} \right)^2 \right] \right\}^{-(\kappa+1)} \qquad (4.2)$$

where $A_\kappa = n_0 \pi^{-3/2} \left( \kappa - 3/2 \right)^{-3/2} v_{th,\perp}^2 v_{th,\parallel} \Gamma(\kappa + 1) \left[ \Gamma(\kappa - 1/2) \right]^{-1}$. For 1 AU parameters, the core is best modelled by a bi-Maxwellian, while the halo and strahl are best modelled by a bi-Kappa as shown in Fig. 3 where the maximum kinetic energy is 1 keV. The following values are from the mean observations in Wilson III et al. (2019). The initial isotropic core has density $n_c = 13.7\,\text{cm}^{-3}$, zero drift, and $v_{th} = v_{th,\parallel} = v_{th,\perp} = 1\,800\,\text{km/s}$. The halo is also isotropic with $n_h = 0.52\,\text{cm}^{-3}$ and $v_{th} = 3\,900\,\text{km/s}$. The strahl has $n_s = 0.21\,\text{cm}^{-3}$, $v_{o,\parallel} = 2\,000\,\text{km/s}$, and $v_{th,\parallel} = 3v_{th,\perp} = 3\,600\,\text{km/s}$. These VDFs are sampled with $\sim 400\,000$ electrons initiated uniformly in speed with pitch angles (the polar angle) from 0 to 180° in increments of 1° and gyrophases (the azimuthal angle) from 0 to 360° in increments of 30°. For 0.3 AU, the core and strahl are modelled with the bi-Maxwellian in parameters similar to Micera et al. (2020), based on observations by Halekas et al. (2020a) (see Fig. 2). The core has $n_c = 332.5\,\text{cm}^{-3}$ and $v_{th} = 3\,900\,\text{km/s}$ with a drift $v_{o,\parallel} = -480\,\text{km/s}$, while the strahl has $n_s = 17.5\,\text{cm}^{-3}$, $v_{th,\parallel} = 7\,900\,\text{km/s}$, $v_{th,\perp} = 5\,600\,\text{km/s}$, and $v_{o,\parallel} = 9\,300\,\text{km/s}$. Since the thermal velocities are at least twice those at 1 AU, we initiate $\sim 1$ million particles up to 2 keV with the same spacing in the solid angle.

## 4.3 Single particle responses and LCE estimation

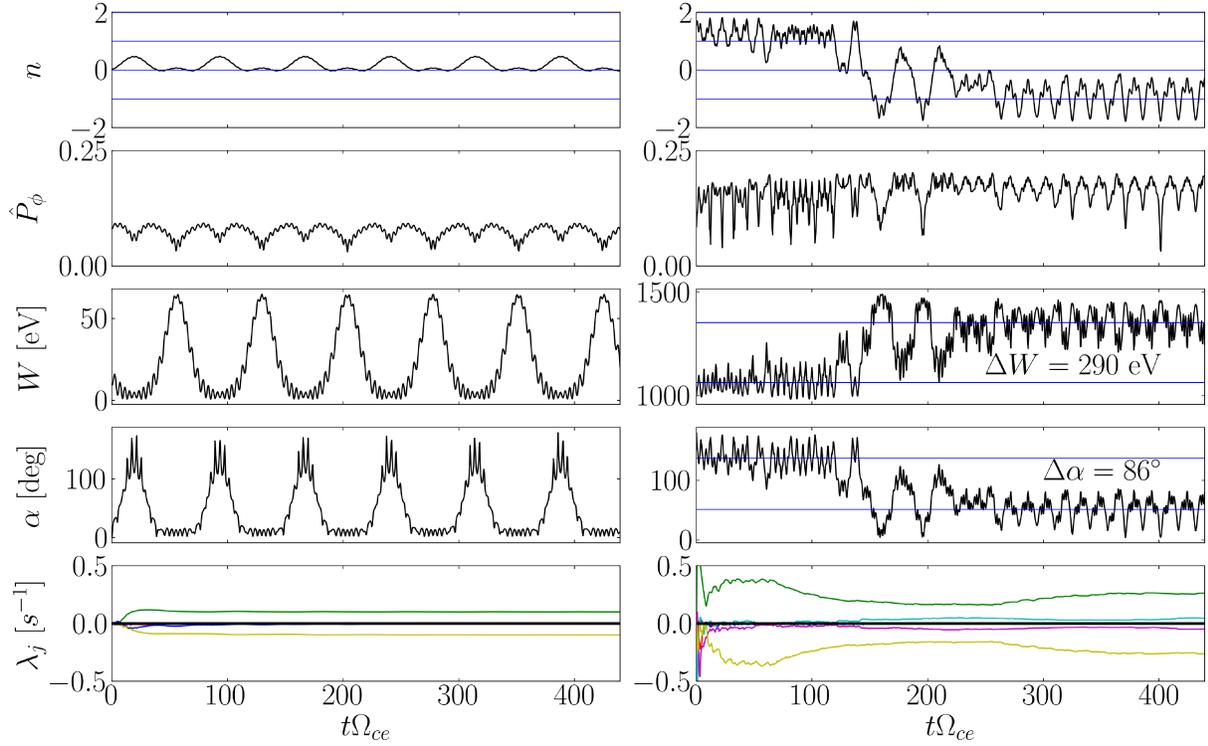

Figure 11: Time series of two electrons with initial conditions $(W_0, \alpha_0) = (10\,\mathrm{eV}, 0°)$ (left column) and $(W_0, \alpha_0) = (1\,\mathrm{keV}, 180°)$ (right column) interacting with a single $65°$ whistler at 1 AU. The first row shows the resonant mismatch $n$ calculated from Eq. (2.13). The second row is the adiabatic invariant $\hat{P}_\phi$ conjugate to the transformed gyrophase $\phi$. The third row is the kinetic energy $W = (\gamma - 1)mc^2$. The fourth row is the pitch angle $\alpha = \cos^{-1}\left(v_z/v\right)$. The last row shows the Lyapunov exponent spectrum $\lambda_j$ in colors, each corresponds to one of the six dimensions in the 6-ball, and the LCE in black.

Fig. 11 shows the dramatic differences in the response of two particles. One is fast ($1\,\mathrm{keV}$) and the other is fairly slow ($10\,\mathrm{eV}$). For the slow particle, we can see quasi-periodic motion where it enters the Landau resonance ($n = 0$) briefly, resulting in an energization in $W$ while the pitch angle $\alpha$ remains constant. Note that the adiabatic invariant $\hat{P}_\phi = P_\phi - n\hat{P}_\zeta$ is modulated by $G_n \propto \rho J_n'(k_\perp \rho)$ for small $P_\parallel$ and $\rho$ near the resonance. So for the slow particle, the fluctuations in $\hat{P}_\phi$ are small ($\sim 0.01$) near $n = 0$. For the $1\,\text{keV}$ particle, the energization and scattering is much more significant. As it flips from $n = 1$ (the fundamental cyclotron resonance) to $n = -1$, the kinetic energy $W$ is increased by 30% of its initial energy and it is scattered by $86°$. It is also worth noticing that the particle sporadically enters and exits a resonance in a short time scale, leading to spikes of the order of 0.1 in the adiabatic invariant.

We know from Section 2.2 that the particle's energy and adiabatic invariant are not conserved when it crosses a resonance. So these conservation laws are momentarily broken. However, in this non-relativistic energy range ($W \sim 1\,\text{keV}$), the resonance crossing occurs frequently and sporadically, resulting in less distinctive changes than an example already shown in Fig. 6, which is typical of wave-particle interactions in the radiation belts. This is due to the small wave fields assumption in Section 2.2. Specifically, it is required that $|qA/mu_{\text{max}}| \ll 1$ for the radiation belts conditions to apply. However, in our simulations, we have shown in Fig. 10 that our particles have maximum velocity $u_{\text{max}}$ such that $|qA/mu_{\text{max}}|$ is $\sim \mathcal{O}(0.1)$. So the simulation of large amplitude whistler waves result in nonlinear effects much different from radiation belts context. For the sake of demonstration, we can reach a comparably similar behavior by simulating relativistic electrons. Fig. 12 shows the distinctive jumps in the resonant harmonic $n$ and the adiabatic invariant $\hat{P}_\phi$ for a $1\,\text{MeV}$ electron under the interaction of the same wave parameters as those in Fig. 11. Note that the trapping occurs both near a resonance and outside a resonance. We have not yet developed a method to identify when this happens for our simulation.

The Lyapunov exponent spectrum, i.e., the different components $\lambda_j$, is plotted in different colors in the last row of Fig. 11. Each of the components does not have any physical significance because the 6-D ball is free to rotate along the particle trajectory in our calculations as described in Section 3.2. But they signify that there is always at least one chaotic component, which corresponds to the sporadic violation of the conservation of the adiabatic invariant inherent in our system. Now, it is their sum, the LCE, that is important. As

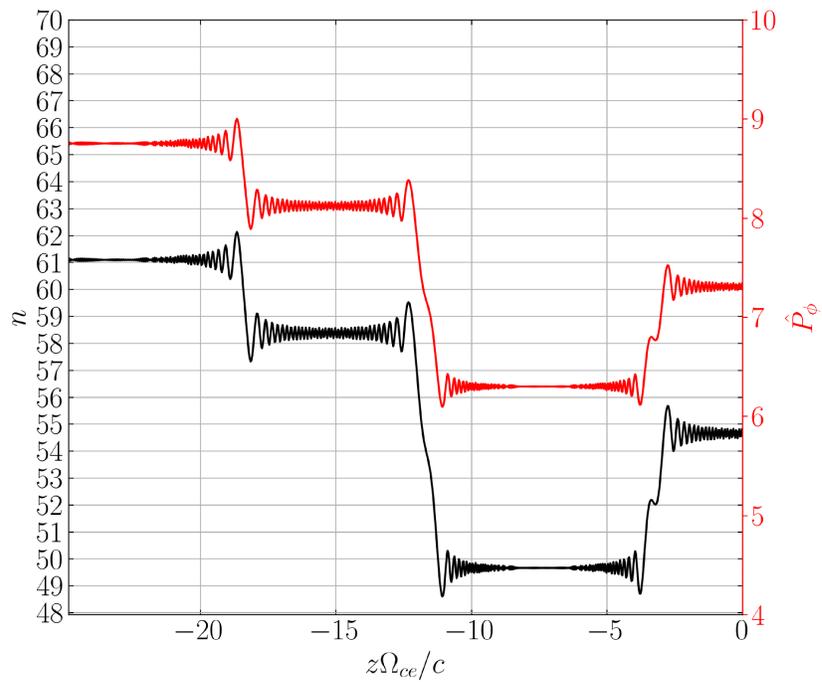

Figure 12: Resonance crossing of a 1 MeV particle interacting with a single 65° whistler at 1 AU. The left (black) axis plots the resonance mismatch $n$, and the right (red) axis shows the adiabatic invariant $\hat{P}_\phi$.

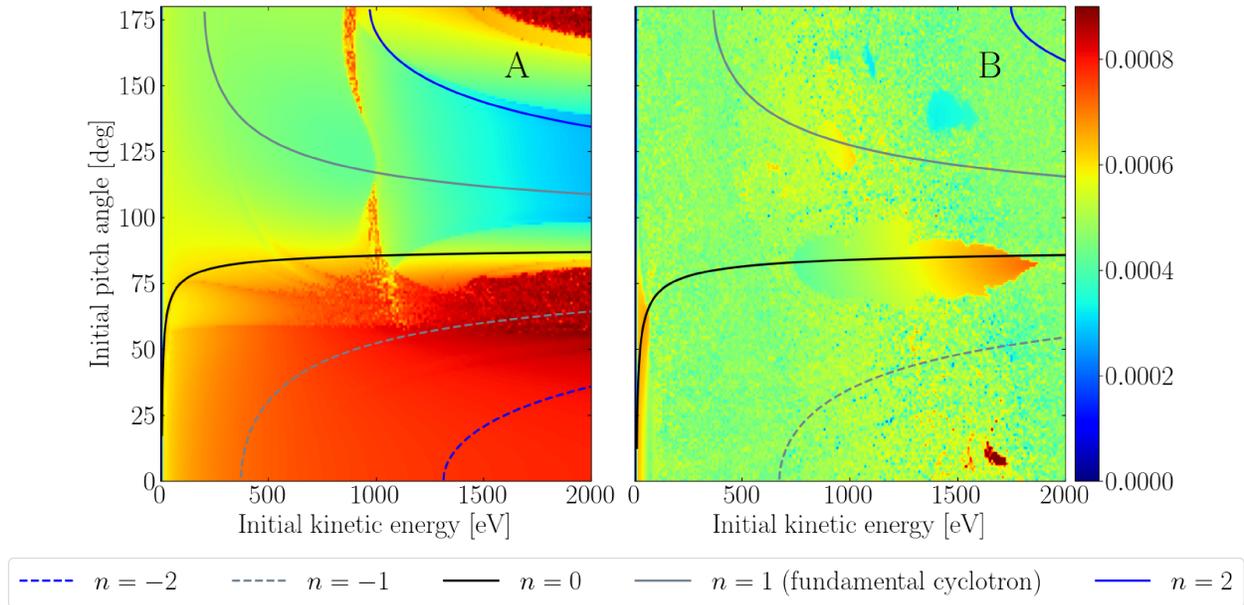

$t f_{ce} = 70.0$

| --- | --- | --- | --- | --- |
| - - - - $n = -2$ | - - - - - $n = -1$ | —— $n = 0$ | —— $n = 1$ (fundamental cyclotron) | —— $n = 2$ |

Figure 13: The LCE of the simulations with a single 5° whistler (A) and a single 65° whistler (B) at 1 AU after a sufficient simulation period for its convergence.

expected, the Lyapunov exponents converge after a transient period at the beginning. So our estimation of the LCE after a sufficiently long time period is constant. This allows us to estimate the best step size to use without running very long simulations. Fig. 13 shows the LCE of all of the initiated particles after it has reached convergence. The maximum LCE is $\lambda \sim 10^{-3}$ in both cases. Thus, for a step size of $\Delta t = 10^{-4}$, the volume of the 6-D ball scales as $\exp(\lambda N \Delta t) \sim 1$ as long as the number of steps $N \lesssim 10^7$. In our physical parameters, $10^7$ steps correspond to $\sim 60$ wave periods, which is a sufficiently long simulation time to study the responses.

## 5 Analysis

In this section, we discuss the structure of the VDF after a long simulation period (60 wave periods for 1 AU and 45 periods for 0.3 AU). In the case of 1 AU parameters, our figures show the core, halo, strahl, and total distribution function, while in 0.3 AU parameters, the distribution includes only the core and strahl electrons. In this non-relativistic range of energy, the $R$ surfaces are almost straight lines, so we only plot the intersections (white crosses) with the $v_\perp = 0$ axis to signify their locations (as derived in Eq. (2.15)). The concentric ellipses (black curves) are the constant $H$ surfaces (from Eq. (2.14)), the center of which is the Landau resonance ($n = 0$). The intersections of the $H$ surfaces with the $v_\perp = 0$ axis show the $n = \pm 1, \pm 2, ...$ radially from the $n = 0$ mode. Recall that in our convention, the $n < 0$ modes are always along the parallel velocity range and the $n > 0$ modes are along the anti-parallel range (as opposite to most papers in the literature).

## 5.1 Single whistlers at 1 AU

For single whistlers, we show the results from three simulations in Fig. 14, which demonstrate the interactions with (from top to bottom) an almost parallel (5°), an almost antiparallel (175°), and an oblique (65°) wave after 60 wave periods. The final VDF of the two parallel cases approximately mirror each other, However, the structures are not entirely identical since the background field points along the wave in one case and against the wave in another, while the strahl electrons are propagating along the field. The first two rows indicate that parallel waves are able to scatter electrons to a certain extent. However, there is a prominent bow-like feature near the $n = -1$ mode at an angle of $\sim 50°$ around the $v_\perp = 0$ axis, which is most apparent for the anti-parallel case. The last row indicates that the interaction with an oblique whistler efficiently isotropizes the strahl, which results in a structure almost identical to the halo by the end of the simulation period. However, there is a lack of high energy and parallel propagating particles, which has been observed in the energy-pitch angle distribution in PSP data (Cattell et al., 2021b). This makes the final results not completely isotropic. Thus, it is not suitable to apply the fitting procedure for the model defined in Eq. (4.2) that Wilson III et al. (2019) used for satellite observations of the VDF.

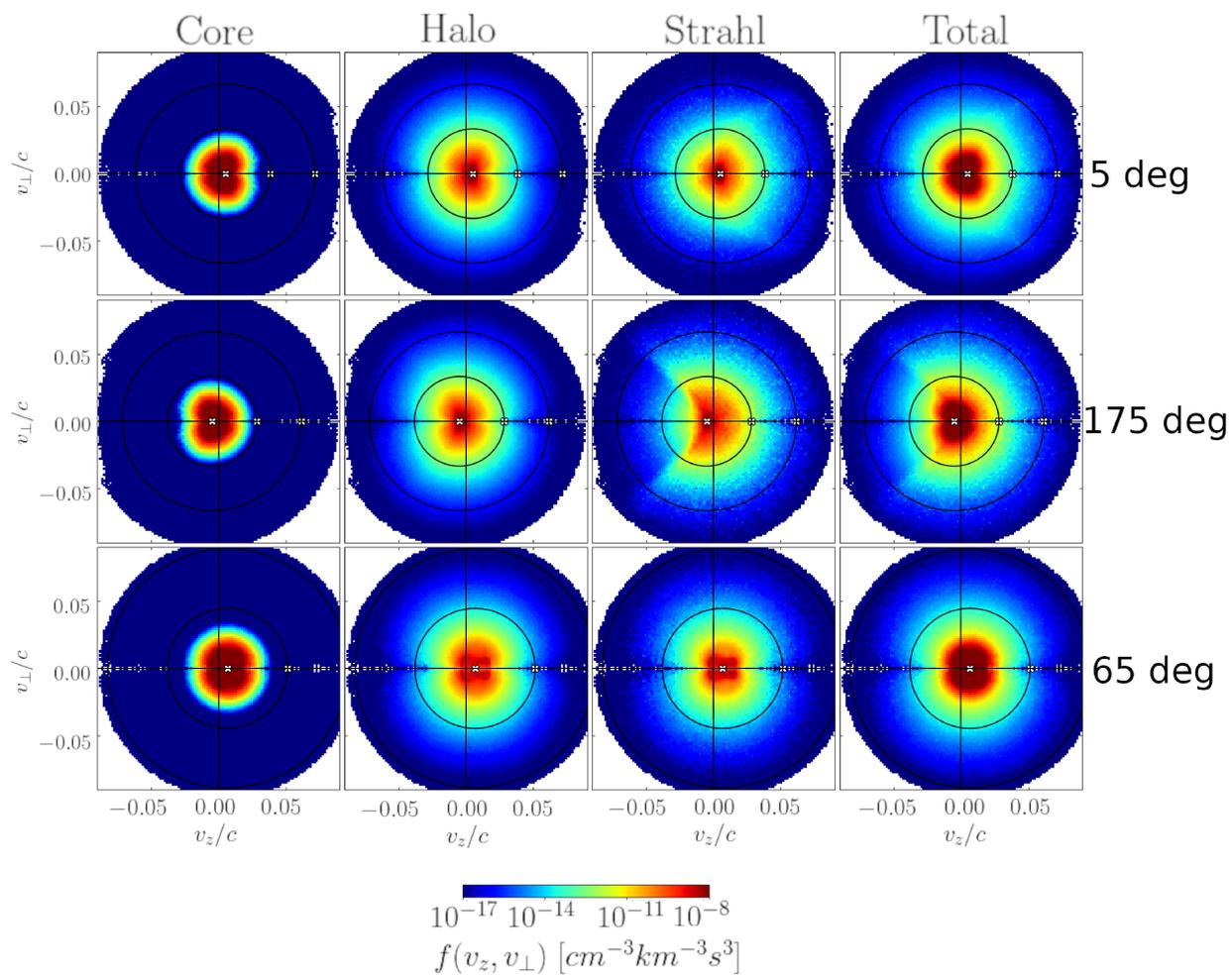

Figure 14: VDF of electron populations after 60 wave periods of interaction with a single whistler at 1 AU. From top to bottom, the rows are the simulations with the 5°, 175°, and 65° wave.

To better understand the VDF structures, we plot the trajectories of a few particles interacting with the 5° and 65° waves during the entire simulation period in Fig. 15. In panels A1, B1, C1, and D1, the particles move along their corresponding $H$ surfaces as expected. The corresponding histograms (A2, B2, C2, and D2) show the points along the particle's trajectory where they hover around the most. In the interaction with the 5° wave (panels A and B), the histograms are uniform, indicating that the particles bounce back and forth in a quasi-periodic motion. There is a point of "reflection" for each energy, which results in the bow-like feature in the VDF. These points are close to the intersections of the $H$ surfaces and the $n < 0$ resonances. This can be due to a combined effect of (a) magnetic mirroring due to the large wave fields comparable to the ambient field and (b) resonant interaction. Effect (a) is a speculation that needs further analysis beyond the scope of this thesis. Here we shall only offer an explanation for (b) from the theory of resonance derived in Section 2.

For $n < 0$, the electron overtakes the wave when it observes a left-hand polarized electromagnetic field in its own frame. Thus, being a right-hand particle, it no longer interacts resonantly. This results in the deceleration of $v_z$ to the negative range where resonant interaction is enabled once again because the particle observes a right-hand polarized wave. It would be interesting to study whether this occurs for a self-consistently simulated wave-particle interaction using PIC code. This is because the $n < 0$ modes are usually where the particle transfers its energy to the wave as it rotates out of phase with the fields, leading to wave generation instead of damping (Tsurutani & Lakhina, 1997). Thus, the wave is modified due to this type of quasi-parallel whistler heat-flux instability (Roberg-Clark et al., 2019; Micera et al., 2020) and both the $H$ and $R$ surfaces are altered accordingly. This might allow the VDF to become more isotropic for particles under interactions with parallel whistlers.

Because the polarization for an oblique wave is elliptical, it is a combination of both right-hand and left-hand waves. This allows for anomalous interactions to occur, which happens when an electron (right-hand) interacts with the $n > 0$ harmonics of a left-hand wave by overtaking it and observing a right-hand polarized electromagnetic field (Tsurutani & Lakhina, 1997). Consequently, the scattering of electrons is more isotropic in panels C and D, where both parallel and anti-parallel particles behave similarly.

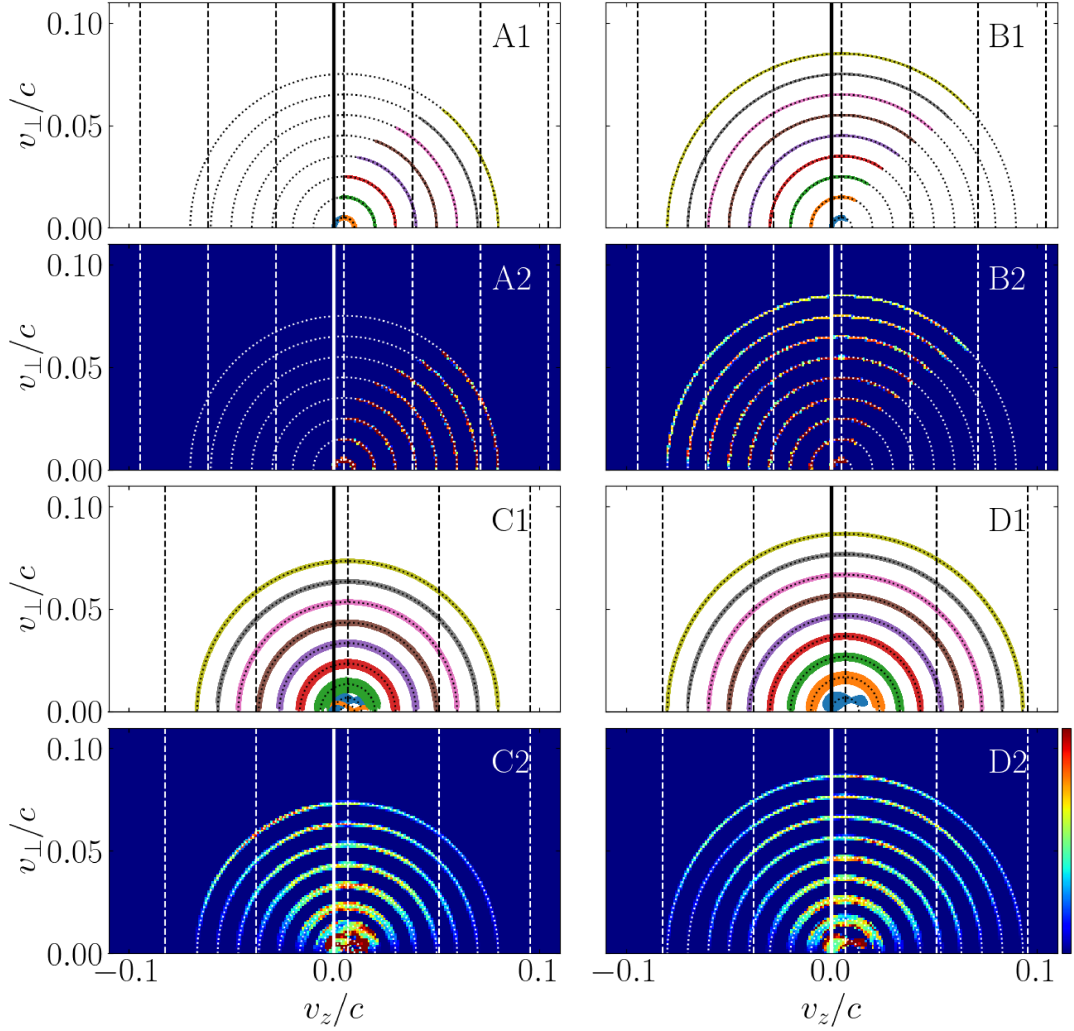

Figure 15: Traces of electron trajectories in the entire simulation period (1 AU parameters) and their corresponding histograms. The panels show those for electrons that are originally parallel (A1-A2) and antiparallel (B1-B2) to the single 5° whistler. Similarly, (C1-C2) and (D1-D2) are those parallel and antiparallel to the single 65° whistler. Their initial speeds are $0, 0.01, 0.02, ..., 0.08c$, which correspond to $0, 26, 102, ...1643\,\mathrm{eV}$. The dotted curves are the constant $H$ ellipses corresponding to the particle's initial energy, while the dashed straight lines are the $R$ surfaces corresponding to (from left to right) $n = 3, 2, ..., -3$. The solid lines are the $v_z = 0$ axis.

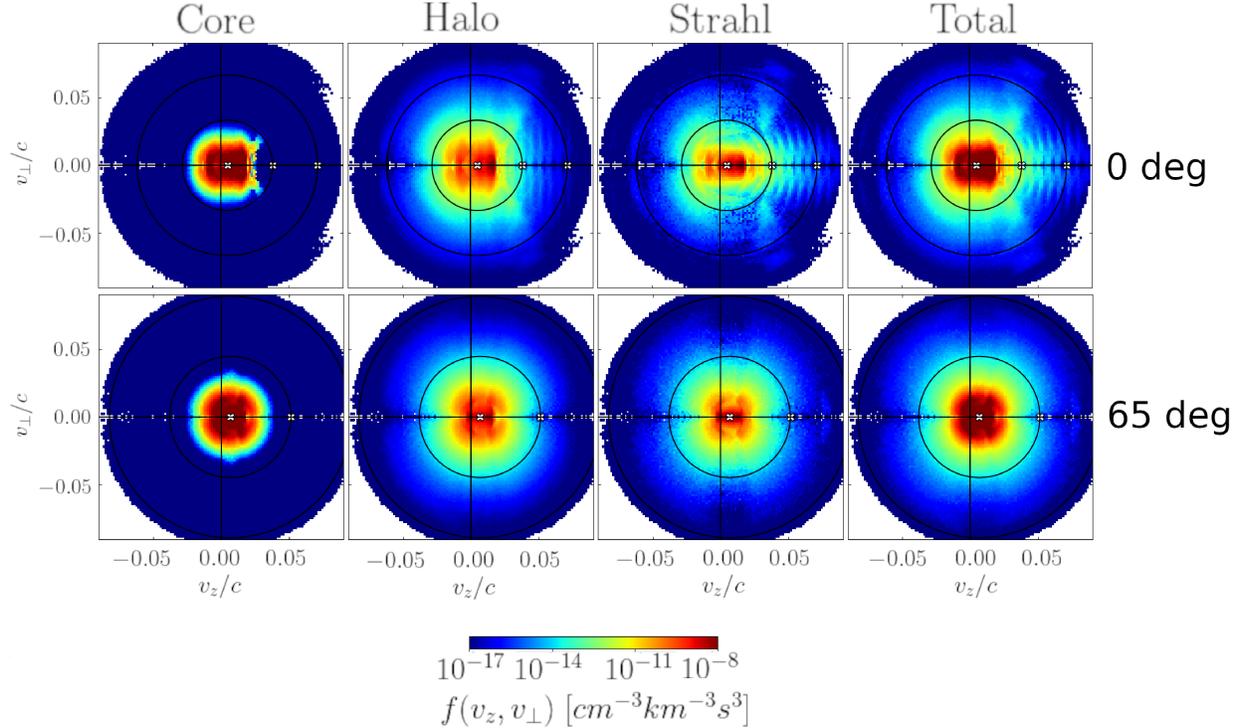

Figure 16: VDF of electron populations after 60 wave periods of interaction with a 0° (top row) and 65° (bottom row) whistler packet at 1 AU.

## 5.2   Whistler packets at 1 AU

Fig. 16 shows the final VDF after 60 wave periods of interaction with a 0° packet and a 65° packet in 1 AU parameters. A region of particle loss similar to that in the single wave parallel whistlers in the previous section can be seen in the 0° packet. However, in addition to the bow-like region, there is also a vertical structure near the $n = -1$ harmonic. Because multiple frequencies are contained in the packet, the $R$ surfaces are now clustered. Fig. 17 shows the surfaces corresponding to the packet's mean frequency wave. In panel B2, there are electrons trapped around the $n = -1$ cluster of $R$ surfaces. Those particles are energetic enough to enter the envelope of the cluster but they cannot escape, resulting in this vertical structure in the VDF. This further supports the explanation from the theory of resonance. In the frame of the mean-frequency wave, there are other waves of different frequencies, which move in both directions with respect to it. Their combined effects cause the trapping around $n = -1$. However, large amplitude waves at $0°$ are not seen in the solar wind at 1 AU. The large amplitude waves at 1 AU are oblique, like the $65°$ packet. For this case, the electron interaction with the packet is similar to the case of a single whistler for the strahl.

The scattering of particles interacting with the oblique packet is highly localized and often in between the resonances (see panels C2 and D2). This is most likely due to the overlapping resonance widths associated with each mode (Karimabadi et al., 1990). Our single-wave resonance surfaces are spaced fairly closely between each harmonic $n$. The overlap of resonance widths can cause more nonlinear and complicated interactions to occur. This topic is beyond the scope of this thesis, so we will not discuss the calculation of the widths.

## 5.3   Whistler packets at 0.3 AU

For interactions with whistler packets in 0.3 AU parameters, we observe the formation of "horn"-like features in the VDF at the locations of the $R$ intersections in the case of oblique propagation (see Fig. 18). This is similar to what was reported in Roberg-Clark et al. (2019). However, they studied very relativistic electrons, which resulted in more defined horn features as the particle velocity term dominates in the canonical momentum. In our parameters, this dominance is weaker, which results in broader horns. Parallel packets do not scatter the strahl as efficiently as oblique packets near the Sun, as similar to the discussion in Vocks et al. (2005). Halekas et al. (2020b) reported that the heat flux observed at 0.3 AU was consistent with the threshold for oblique whistler fan instability. So our results are consistent with near-Sun observations.

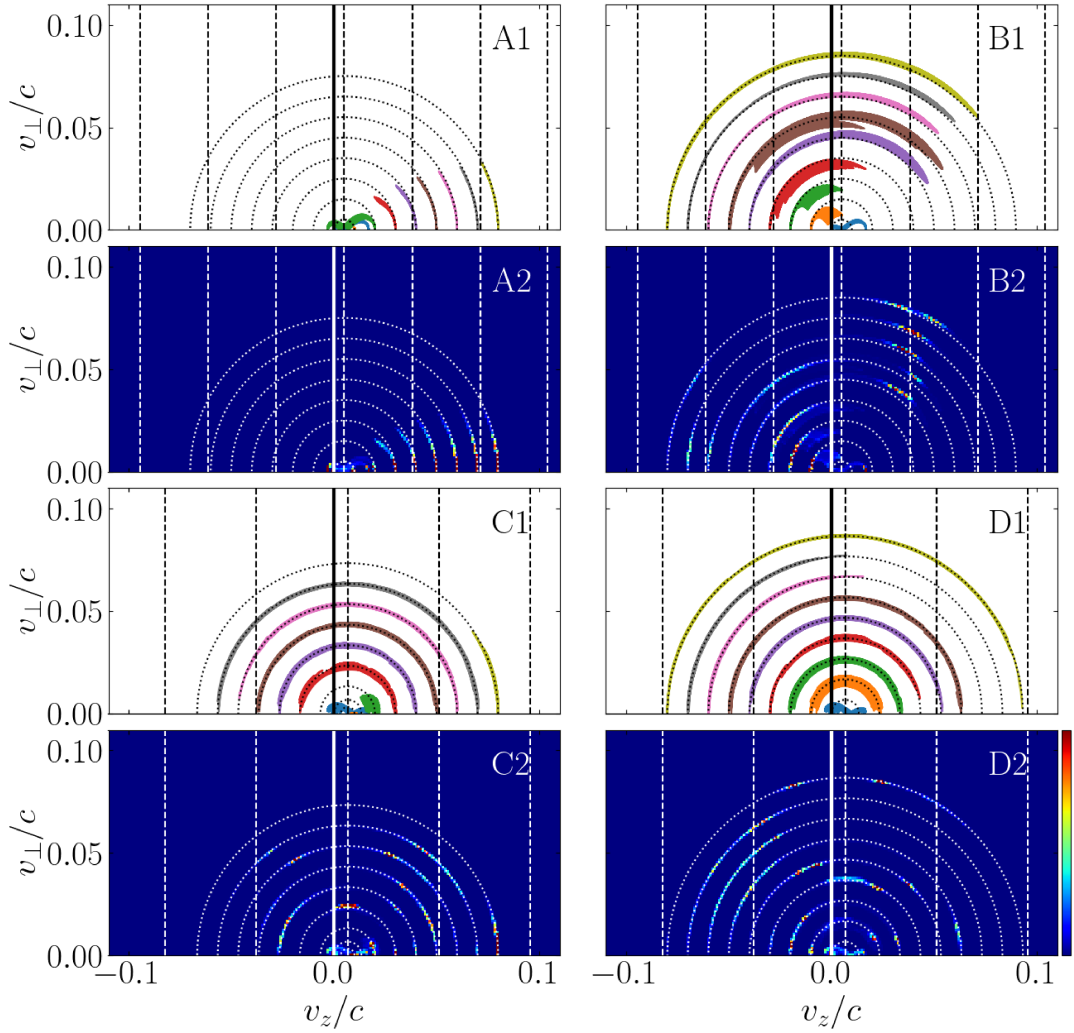

Figure 17: Traces of electron trajectories in the entire simulation period (1 AU parameters) and their corresponding histograms. The panels are similar to those in Fig. 15, but the electrons are under interactions with a 0° packet (A and B) and a 65° packet (C and D).

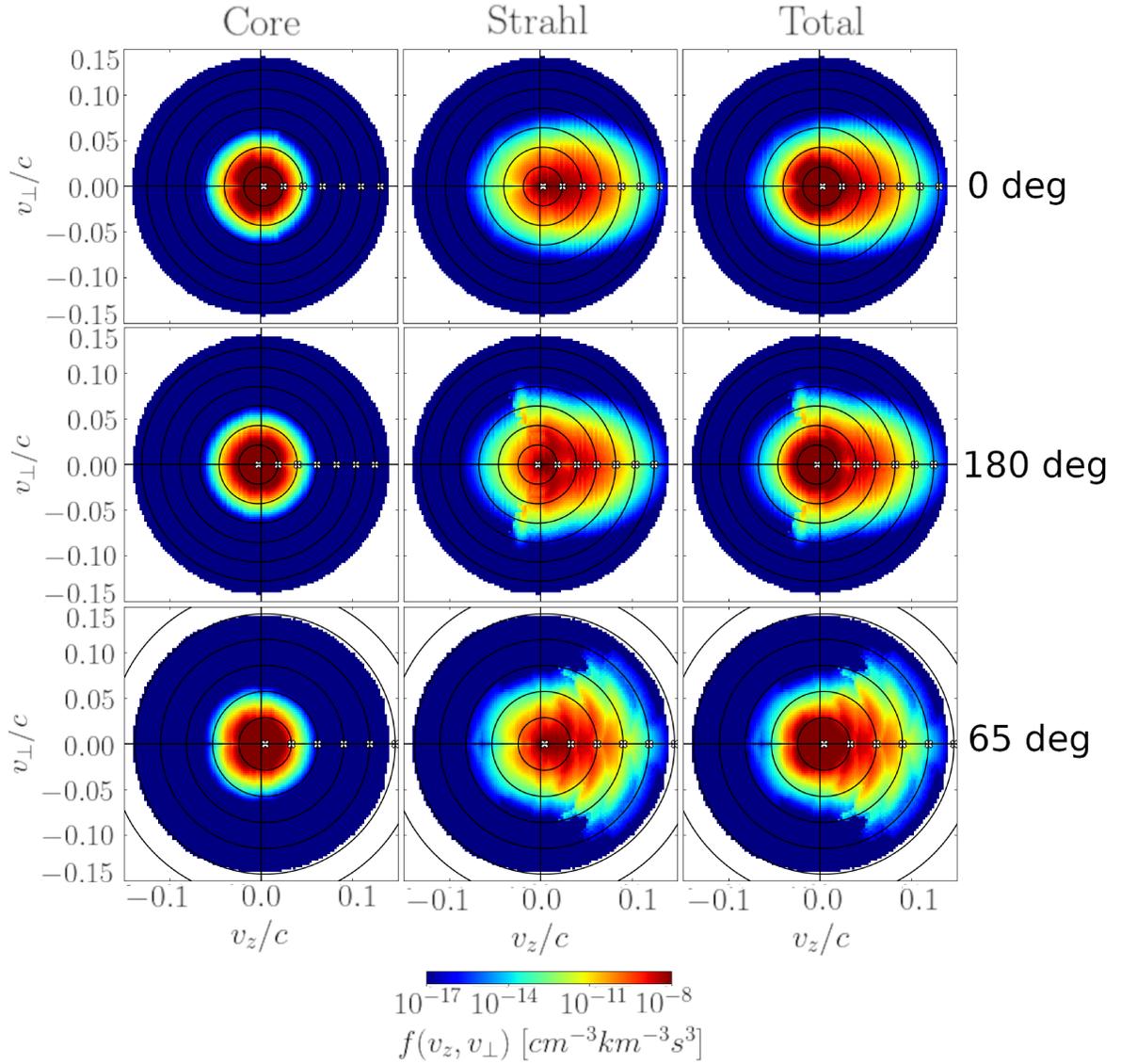

Figure 18: VDF of electron populations after 45 wave periods of interaction with whistler packets at 0.3 AU. From top to bottom, the rows are the simulations with the 0°, 180°, and 65° wave.

# 6 Conclusion

We have used a vectorized test particle simulation to study the scattering and energization of solar wind electrons from their interactions with single whistlers and whistler packets at different propagation angles and in 0.3 AU and 1 AU background parameters. We showed that for non-relativistic particles, the interaction is mainly a diffusion in pitch angle. The particles are scattered along the constant $H$ surface, while interacting with the nearest resonant mode. Our results show that the final velocity distribution function at 0.3 AU are consistent with observations from PSP (Cattell et al., 2021b; Halekas et al., 2020b) and with simulations in Roberg-Clark et al. (2019) and Micera et al. (2020) for the case of obliquely propagating whistler packets. Resonant strahl electrons are scattered to higher pitch angle, until they can be characterized as an isotropic halo. This verifies the theory that the origin of the halo is the strahl, since these waves can scatter the VDF in a short length scale. Thus, it explains the existence of a halo population of electrons far from the Sun at 1 AU and the corresponding heliospheric radial decrease in strahl density. We also observed that parallel waves are less efficient in isotropizing the electron distribution, consistent with the heat-flux study in Halekas et al. (2020b).

# 7 Future works

Much of the analysis can be further extended from the basis laid out in this thesis using the Hamiltonian approach. The resonance widths of the harmonics can be calculated to determine the overlapping and their subsequent effects on the VDF structure. Since the derived adiabatic invariants are analogous to the magnetic moment for a system without the wave perturbations, they can be used to determine the constraints on the velocity, which describes the magnetic mirroring effect. More interesting physics might be revealed by simulations of the interaction of relativistic electrons with the large amplitude waves described in this thesis. This is because the small field assumptions of $\delta_{1,2}$ are better satisfied if the particle momentum term dominates in the canonical momentum. In terms of the simulation program, our code is written purely in Python and is vectorized with Python arrays, but better optimization can be achieved with true SIMD vectorization in C. This simulation program might be further developed into a full PIC code with the implementation of field solving components, in which case, a translation to C is absolutely necessary.

# Appendices

## A  Useful mathematical theorems

**Theorem A.1** (Leibniz's integral rule)**.** Given a bounded domain $I \subseteq \mathbb{R}$ with bounded boundaries $a(x), b(x)$ defined on $I$, let $f(x,t)$ be a $C^1$ smooth function on $I \times \big[a(x), b(x)\big]$. Also, suppose $a, b$ are $C^1$ smooth. Then for $x \in I$,

$$\frac{d}{dx}\left(\int_{a(x)}^{b(x)} f(x,t)dt\right) = f(x, b(x)) \cdot \frac{db(x)}{dx} - f(x, a(x)) \cdot \frac{da(x)}{dx} + \int_{a(x)}^{b(x)} \frac{\partial f(x,t)}{\partial x}dt \quad \text{(A.1)}$$

**Theorem A.2** (Bessel decomposition)**.** For $x \in \mathbb{R}^+$ and $\phi, \delta \in \mathbb{R}$, the following sinusoidal functions with oscillatory phase can be decomposed into Bessel-Fourier series

$$\sin\left(x \sin\phi + \delta\right) = \sum_{n \in \mathbb{Z}} J_n(x) \sin\left(n\phi + \delta\right) \quad \text{(A.2a)}$$

$$\cos\phi \sin\left(x \sin\phi + \delta\right) = \sum_{n \in \mathbb{Z}} \frac{n}{x} J_n(x) \sin\left(n\phi + \delta\right) \quad \text{(A.2b)}$$

$$\sin\phi \cos\left(x \sin\phi + \delta\right) = \sum_{n \in \mathbb{Z}} J_n'(x) \sin\left(n\phi + \delta\right) \quad \text{(A.2c)}$$

where $J_n(x)$ are the $n$th order Bessel functions of the first kind and $J_n'(x)$ are their first order derivative.

## B  The Hamiltonian resonance analysis

In this section, we perform two transformations to reduce the Hamiltonian in Eq. (2.8) into an integrable 1-D form, from which the adiabatic invariants can be calculated and the resonant condition is derived from the equation of motion. Assuming that the wave fields are small, we can write $\mathcal{H} = \mathcal{H}_0 + \mathcal{H}_1$ in a power series of $A_1$ and $A_2$ to the first order as follows.

$$\mathcal{H} = \gamma mc^2 - \frac{qA_1}{\gamma m}\left[\left(\frac{k_\parallel}{k}\right)P_x - \left(\frac{k_\perp}{k}\right)P_z\right]\sin\psi - \frac{qA_2}{\gamma m}\left(P_y - qB_0 x\right)\cos\psi + q\Phi_0\sin\psi \quad \text{(B.1)}$$

where $\mathcal{H}_0 = \gamma mc^2$ is the Hamiltonian without the presence of any wave and

$$\gamma = \sqrt{1 + \frac{P_x^2}{m^2c^2} + \frac{(P_y - qB_0 x)^2}{m^2c^2} + \frac{P_z^2}{m^2c^2}} \quad \text{(B.2)}$$

is the Lorentz factor. From $\mathcal{H}_0$, we can invert and solve for $P_x$

$$P_x = |q|B_0\sqrt{\frac{\mathcal{H}_0^2/c^2 - m^2c^2 - P_z^2}{q^2B_0^2} - (x - X)^2} = |q|B_0\sqrt{\rho^2 - (x - X)^2} \quad \text{(B.3)}$$

where we have written $X = P_y/qB_0$ and $\rho$ such that $\mathcal{H}_0 = \sqrt{m^2c^4 + q^2B_0^2\rho^2c^2 + P_z^2c^2}$. Now, define the action $J = (1/2\pi)\oint P_x dx = |q|B_0\rho^2/2$, which is just the area of a circle with radius $\rho$ centered at $x = X$. Let $2|q|B_0 J = P_\perp^2$ and $P_z = P_\parallel$. The Hamiltonian then becomes

$$\mathcal{H}_0 = \sqrt{m^2c^4 + 2|q|B_0 Jc^2 + P_\parallel^2 c^2} = \sqrt{m^2c^4 + P_\perp^2 c^2 + P_\parallel^2 c^2} \quad \text{(B.4)}$$

$J$ is thus analogous to the perpendicular momentum and we can interpret $\rho = P_\perp/|q|B_0 = v_\perp/\Omega_c$ as the particle's gyroradius where $\Omega_c = |q|B_0/m$ is the cyclotron frequency. Eq. (B.4) is now independent of the conjugate coordinates. Thus, if we define the new momentum as $P_\phi = J$, then it is an adiabatic invariant.

Now, we need to find the coordinate conjugate to $P_\phi$. Define the generating function $F_2(x, y, z; P_\phi', P_y', P_\parallel') = \int_X^x d\overline{x}P_x(\overline{x}; P_\phi', P_y') + yP_y' + zP_\parallel'$ where new variables are denoted with a prime. The old momenta transform trivially as

$$P_x = \frac{\partial F_2}{\partial x} = P_x, \qquad P_y = \frac{\partial F_2}{\partial y} = P_y', \qquad P_\parallel = \frac{\partial F_2}{\partial z} = P_\parallel' \quad \text{(B.5)}$$

where the first equation is true due to the First Fundamental Theorem of Calculus. The new $z$ coordinate can also be found easily $z' = \partial F_2 / \partial P'_\parallel = z$. Now, the coordinate conjugate to $P'_\phi$ is given by

$$\phi' = \frac{\partial F_2}{\partial P'_\phi} = \int_X^x \frac{d\overline{x}}{\sqrt{\rho^2 - (\overline{x} - X)^2}} = \sin^{-1}\left(\frac{x - X}{\rho}\right) \tag{B.6}$$

Inverting, we get $x = X + \rho \sin \phi'$. This simplifies the momentum $P_x$ in Eq. (B.3) to $P_x = |q| B_0 \rho \cos \phi'$. Similarly, the new $y$ coordinate is

$$y' = \frac{\partial F_2}{\partial P'_y} = y + \frac{1}{qB_0} \frac{\partial F_2}{\partial X} = y + s \frac{\partial}{\partial X} \int_X^x d\overline{x} \sqrt{\rho^2 - (\overline{x} - X)^2} \tag{B.7}$$

Note that both the integrand and the boundary are dependent on $X$. So we have to use Theorem A.1

$$\frac{\partial}{\partial X} \int_X^x d\overline{x} \sqrt{\rho^2 - (\overline{x} - X)^2} = -\rho + \int_X^x d\overline{x} \frac{\overline{x} - X}{\sqrt{\rho^2 - (\overline{x} - X)^2}} = -\rho \cos \phi' \tag{B.8}$$

where $\phi'$ is found in Eq. (B.6). We can then invert to find $y = y' + s\rho \cos \phi'$. Now, we can drop the prime and write the transformed Hamiltonian as $\mathcal{H}_0 = \sqrt{m^2 c^4 + 2|q| B_0 P_\phi c^2 + P_\parallel^2 c^2}$.

Under this transformation, the perturbation $\mathcal{H}_1$ become

$$\mathcal{H}_1 = q \left[ \Phi_0 + \frac{A_1}{\gamma} \left(\frac{k_\perp}{k}\right) \frac{P_\parallel}{m} \right] \sin \psi - \frac{qA_1}{\gamma} \left(\frac{k_\parallel}{k}\right) \rho \Omega_c \cos \phi \sin \psi + s \frac{qA_2}{\gamma} \rho \Omega_c \sin \phi \cos \psi \tag{B.9}$$

where the new wave phase is $\psi = k_\perp \rho \sin \phi + k_\perp X + k_\parallel z - \omega t$. From Theorem A.2, this Hamiltonian can be written in the form $\mathcal{H}_1 = \sum_{n \in \mathbb{Z}} G_n(P_\phi, P_\parallel) \sin \zeta_n$ where

$$G_n(P_\phi, P_\parallel) = q \left[ \Phi_0 + \frac{A_1}{\gamma} \left(\frac{k_\perp}{k} \frac{P_\parallel}{m} - \frac{k_\parallel}{k} \frac{n\Omega_c}{k_\perp}\right) \right] J_n(k_\perp \rho) + s \frac{qA_2}{\gamma} \rho \Omega_c J'_n(k_\perp \rho) \tag{B.10}$$

and the phase is now $\zeta_n = n\phi + k_\perp X + k_\parallel z - \omega t$. Its time derivative is $\dot{\zeta}_n = -\omega + n\dot{\phi} + k_\parallel \dot{z}$.

Using the equation of motion from $\mathcal{H}_0$ and $\mathcal{H}_1$, we can write

$$\frac{d\zeta_n}{dt} = F_n + \sum_{l \in \mathbb{Z}} \left( l \frac{\partial G_l}{\partial P_\phi} + k_\parallel \frac{\partial G_l}{\partial P_\parallel} \right) \sin \zeta_l = F_n + \sum_{l \in \mathbb{Z}} \mathcal{F}_l \sin \zeta_l \qquad \text{(B.11)}$$

where $F_n = -\omega + n \partial \mathcal{H}_0 / \partial P_\phi + k_\parallel \partial \mathcal{H}_0 / \partial P_\parallel$. Recall that we have assumed small wave fields, i.e. $|\mathcal{F}_l| \ll |F_n|$. So the motion of $\zeta_n$ is usually fast $(|\dot{\zeta_n}| > \Omega_c)$ with small oscillations around $F_n$. However, whenever $|F_n| \to 0$, this no longer holds, the particle comes into resonance with the $l = n$ mode, and $\zeta_n$ is slowly varying. Thus, $F_n$ determines the resonant condition. To analyze this in more detail, we isolate the $l = n$ mode from the infinite series since it contributes more significantly than other terms (Albert, 2000). To transform into the wave frame, we use the generating function $F_2(x, y, z; \hat{P}_\phi, \hat{P}_y, \hat{P}_\zeta; t) = \phi \hat{P}_\phi + y \hat{P}_y + \left( n\phi + k_\perp X + k_\parallel z - \omega t \right) \hat{P}_\zeta$. The new variables obey the following relations

$$\hat{\phi} = \phi \qquad\qquad \hat{y} = y + k_\perp \hat{P}_\parallel / q B_0 \qquad\qquad \zeta = n\phi + k_\perp X + k_\parallel z - \omega t$$

$$\hat{P}_\phi = P_\phi - n P_\parallel / k_\parallel \qquad \hat{P}_y = P_y \qquad\qquad \hat{P}_\zeta = P_\parallel / k_\parallel \qquad\qquad \text{(B.12)}$$

where we have omitted the subscript $n$ in $\zeta$ for simplicity. The total Hamiltonian becomes

$$\mathcal{H}(\zeta; \hat{P}_\phi, \hat{P}_\zeta) = \mathcal{H}_0(\hat{P}_\phi + n\hat{P}_\zeta, k_\parallel \hat{P}_\zeta) - \omega \hat{P}_\zeta + G_n(\hat{P}_\phi + n\hat{P}_\zeta, k_\parallel \hat{P}_\zeta) \sin \zeta \qquad \text{(B.13)}$$

where $\hat{P}_\phi$ is an adiabatic invariant. Also, $d\hat{P}_\zeta / dt = -\partial \mathcal{H} / \partial \zeta = -G_n \cos \zeta$. When the phase varies rapidly, we can average $\zeta$ over its period and find $d\hat{P}_\zeta / dt = 0$ since $\langle \cos \zeta \rangle = 0$. The momentum $\hat{P}_\zeta$ is then an adiabatic invariant. However, when $\dot{\zeta}$ is slow, $P_\zeta$ is no longer conserved and undergoes irreversible changes. As mentioned above, this happens when

$$-\omega + n \frac{\partial \mathcal{H}_0}{\partial P_\phi} + k_\parallel \frac{\partial \mathcal{H}_0}{\partial P_\parallel} = -\omega + \frac{n\Omega_c}{\gamma} + \frac{k_\parallel P_\parallel}{\gamma m} = 0 \qquad\qquad \text{(B.14)}$$

# C   Constant energy and resonant surfaces

In this section, we lay out the mathematical steps to derive Eq. (2.14) and Eq. (2.15). The latter is easier to derive. The Lorentz factor at $v_\perp = 0$ is $\gamma = \left(1 - v_\parallel^2/c^2\right)^{-1/2}$. So, multiplying the resonant condition in Eq. (2.13) by $1/ck_\parallel$ and reorganizing yield

$$\left(1 + \alpha_n^2\right)\left(\frac{v_\parallel}{c}\right)^2 - 2v_p\frac{v_\parallel}{c} + v_p^2 - \alpha_n^2 = 0 \qquad (C.1)$$

where $\alpha = n\Omega_c/k_\parallel c$ and $v_p = \omega/k_\parallel c$. The roots of this quadratic equation in $v_\parallel/c$ are Eq. (2.15).

Now, we focus on Eq. (2.14). Let $H_0 = \gamma - v_p(P_\parallel/mc)$ be a constant. Then we can invert it into the following form as given in Karimabadi et al. (1990),

$$a_H\left(\frac{P_\parallel}{mc} + \frac{b_H}{a_H}\right)^2 - \frac{P_\perp^2}{m^2c^2} = \frac{b_H^2}{a_H} + 1 - H_0^2 \qquad (C.2)$$

where $a_H = v_p^2 - 1$ and $b_H = v_pH_0$. First, the Lorentz factor can be written as $\gamma = \left(1 + P_\perp^2/m^2c^2 + P_\parallel^2/m^2c^2\right)^{1/2}$. Expanding the equation above and simplifying yield

$$\left(H_0 + v_p\frac{P_\parallel}{mc}\right)^2 = \gamma^2 \qquad (C.3)$$

Taking the square root and letting $P_\parallel/mc = \gamma v_\parallel/c$, this becomes $H_0/\gamma = 1 - v_pv_\parallel/c$. Now, we write $\gamma = \left(1 - v_\perp^2/c^2 - v_\parallel^2/c^2\right)^{-1/2}$. Squaring and simplifying yields

$$\left(H_0^2 + v_p^2\right)\left(\frac{v_\parallel}{c} - \frac{v_p}{H_0^2 + v_p^2}\right)^2 + H_0^2\frac{v_\perp^2}{c^2} = H_0^2 - 1 + \frac{v_p^2}{H_0^2 + v_p^2} = R_0 \qquad (C.4)$$

This is the same as Eq. (2.14). So we are done.

# D   The local expansion operator

The non-relativistic equation of motion can be written as $d\mathbf{p}/dt = \mathbf{F}(t, \mathbf{r}, \mathbf{v})$ where $\mathbf{p} = (\mathbf{r}, \mathbf{v})$ is a position in 6-D phase space and $\mathbf{F}$ is determined by the usual Lorentz equation with $d\mathbf{r}/dt = \mathbf{v}$. It is convenient to expand the wave fields to the complex field here when taking the Jacobian. We write $\mathbf{E} = \left( E_x^w \hat{\mathbf{x}} + i E_y^w \hat{\mathbf{y}} + E_z^w \hat{\mathbf{z}} \right) e^{i\psi}$ and $\mathbf{B} = \left( -i B_x^w \hat{\mathbf{x}} + B_y^w \hat{\mathbf{y}} - i B_z^w \hat{\mathbf{z}} \right) e^{i\psi} + B_0 \hat{\mathbf{z}}$. Then the Jacobian of the system is

$$\boldsymbol{\nabla} \mathbf{F} = \mathbf{F} \otimes \begin{bmatrix} \boldsymbol{\nabla}_{\mathbf{r}} & \boldsymbol{\nabla}_{\mathbf{v}} \end{bmatrix}^T = \begin{pmatrix} 0 & \mathbb{1}_3 \\ (\mathbf{E} + \mathbf{v} \times \mathbf{B}) \otimes (i\mathbf{k}) & \Omega_B \end{pmatrix} \tag{D.1}$$

where $\otimes$ is the outer product, $\mathbb{1}_3$ is the three-dimensional identity matrix, and the magnetic rotation term $\Omega_B = \boldsymbol{\nabla}_{\mathbf{v}}(\mathbf{v} \times \mathbf{B})$ is

$$\Omega_B = \begin{pmatrix} 0 & B_z & -B_y \\ -B_z & 0 & B_x \\ B_y & -B_x & 0 \end{pmatrix} \tag{D.2}$$

Now, consider two "very close" solutions of the equation of motion $\mathbf{X}, \tilde{\mathbf{X}}$ up to a small radius $\delta$. We can write

$$\frac{d(\mathbf{X}(t) - \tilde{\mathbf{X}}(t))}{dt} = \mathbf{F}(t, \mathbf{X}(t)) - \mathbf{F}(t, \tilde{\mathbf{X}}(t)) \sim \boldsymbol{\nabla} \mathbf{F} \cdot \left( \mathbf{X}(t) - \tilde{\mathbf{X}}(t) \right) \tag{D.3}$$

up to a linear approximation. The corresponding forward time, finite difference expression of this is $\Delta \mathbf{X}(t + \Delta t) \approx \Delta \mathbf{X}(t) + \Delta t \boldsymbol{\nabla} \mathbf{F} \cdot \Delta \mathbf{X}(t)$ where $\Delta \mathbf{X} = \mathbf{X} - \tilde{\mathbf{X}}$. Then the local expansion at the $n$th time step can be written as $\mathbf{M}_n = \mathbb{1}_6 + \Delta t \boldsymbol{\nabla} \mathbf{F}_n$ where

$$\Delta \mathbf{X}_{n+1} = \mathbf{M}_n \cdot \Delta \mathbf{X}_n \tag{D.4}$$